\documentclass[usegraphicx,usenatbib,useapjfonts,apj]{emulateapj}

\usepackage{apjfonts}
\usepackage{amsbsy}
\usepackage{graphicx}
\usepackage{dcolumn}

\renewcommand\({\left(}
\renewcommand\){\right)}
\renewcommand\[{\left[}
\renewcommand\]{\right]}

\newcommand{\ra}{\rightarrow}

\def\lsim{\raise 0.4ex\hbox{$<$}\kern -0.8em\lower 0.62
ex\hbox{$\sim$}}

\def\gsim{\raise 0.4ex\hbox{$>$}\kern -0.7em\lower 0.62
ex\hbox{$\sim$}}

\def\lbar{{\hbox{$\lambda$}\kern -0.7em\raise 0.6ex
\hbox{$-$}}}

\newcommand\eq[1]{eq.~(\ref{#1})}
\newcommand\eqs[2]{eqs.~(\ref{#1}) and (\ref{#2})}
\newcommand\Eq[1]{Equation~(\ref{#1})}
\newcommand\Eqs[2]{Equations~(\ref{#1}) and (\ref{#2})}

\newcommand\eqss[3]{eqs.~(\ref{#1}), (\ref{#2}) and (\ref{#3})}

\newcommand\eqst[2]{eqs.~(\ref{#1})--(\ref{#2})}
\newcommand\pa{\partial}
\newcommand\p{\partial}

\newcommand\ee{\end{equation}}
\newcommand\be{\begin{equation}}
\def\bea{\begin{array}}
\def\eea{\end{array}}\def\ea{\end{array}}
\newcommand\ees{\end{eqnarray}}
\newcommand\bees{\begin{eqnarray}}

\def\p1{{\bf p}_1}
\def\p2{{\bf p}_2}
\def\k1{{\bf k}_1}
\def\k2{{\bf k}_2}

\newcommand{\dddM}{\kern 0.2em \raise 1.9ex\hbox{$...$}\kern -1.0em \hbox{$M$}}
\newcommand{\dddQ}{\kern 0.2em \raise 1.9ex\hbox{$...$}\kern -1.0em \hbox{$Q$}}
\newcommand{\dddI}{\kern 0.2em \raise 1.9ex\hbox{$...$}\kern -1.0em\hbox{$I$}}
\newcommand{\dddJ}{\kern 0.2em \raise 1.9ex\hbox{$...$}\kern-1.0em
\hbox{$J$}}
\newcommand{\dddcalJ}{\kern 0.2em \raise 1.9ex\hbox{$...$}\kern-1.0em
\hbox{${\cal J}$}}

\newcommand{\dddO}{\kern 0.2em \raise 1.9ex\hbox{$...$}\kern -1.0em
\hbox{${\cal O}$}}
\def\dddz{\raise 1.5ex\hbox{$...$}\kern -0.8em \hbox{$z$}}
\def\dddd{\raise 1.8ex\hbox{$...$}\kern -0.8em \hbox{$d$}}
\def\dddbd{\raise 1.8ex\hbox{$...$}\kern -0.8em \hbox{${\bf d}$}}
\def\ddbd{\raise 1.8ex\hbox{$..$}\kern -0.8em \hbox{${\bf d}$}}
\def\dddx{\raise 1.6ex\hbox{$...$}\kern -0.8em \hbox{$x$}}

\def\D{\Delta}
\def\p{\partial}
\def\a{\alpha}

\def\nn{\nonumber}
\def\th{\theta}
\def\s{\sigma}
\def\g{\gamma}
\def\G{\Gamma}
\def\d{\delta}

\def\eps{\epsilon}

\def\dslash{\hspace{-1mm}\not{\hbox{\kern-2pt $\partial$}}}
\def\Dslash{\not{\hbox{\kern-4pt $D$}}}
\def\pslash{\not{\hbox{\kern-2.1pt $p$}}}
\def\kslash{\not{\hbox{\kern-2.3pt $k$}}}
\def\qslash{\not{\hbox{\kern-2.3pt $q$}}}

\newcommand{\inT}{\int_{-\infty}^{\infty}}

\newcommand{\Dl}{\int{\cal D}\lambda}

\begin{document}

\title{The Halo Mass Function from  Excursion 
Set Theory.\\
I. Gaussian fluctuations with non-markovian  dependence on the smoothing scale}

\author{
Michele Maggiore\altaffilmark{1} and
Antonio Riotto\altaffilmark{2,3}
}
\altaffiltext{1}{D\'epartement de Physique Th\'eorique, 
Universit\'e de Gen\`eve, 24 quai Ansermet, CH-1211 Gen\`eve, Switzerland}
\altaffiltext{2}{CERN, PH-TH Division, CH-1211, Gen\`eve 23,  Switzerland}
\altaffiltext{3}{INFN, Sezione di Padova, Via Marzolo 8,
I-35131 Padua, Italy}

\begin{abstract}

A classic method for computing 
the mass function  of dark matter halos
is provided by excursion set theory, where  density
perturbations evolve stochastically 
with the smoothing scale,
and the problem of computing the probability of halo formation is
mapped into the so-called first-passage time problem in the presence
of a barrier.  While the full dynamical complexity  of halo
formation can only be revealed through $N$-body simulations, 
excursion set theory provides a simple analytic framework for understanding 
various aspects of this
complex process. In this series of paper we propose improvements of both
technical and conceptual aspects of excursion set theory, and we
explore up to which point the method can reproduce quantitatively
the data from $N$-body simulations. In  paper I of the series we show how to
derive excursion set theory from a path integral formulation. This
allows us both to derive rigorously the absorbing barrier boundary
condition, that in the usual formulation is just postulated, and to 
deal analytically with the non-markovian nature of the random walk.  Such a non-markovian dynamics 
inevitably enters when either the density is smoothed with filters such
as the top-hat filter in coordinate space (which is the only filter
associated to a well defined halo mass) or when one considers
non-Gaussian fluctuations. In these cases,
beside ``markovian'' terms, we find
``memory'' terms that reflect the non-markovianity of the evolution
with the smoothing scale.
We  develop a general 
formalism for evaluating perturbatively these non-markovian
corrections, and in this paper we perform explicitly the computation of
the halo  mass function for gaussian fluctuations, 
to first order in the non-markovian
corrections due to the use of a tophat
filter in coordinate space.

In  paper II of this series we propose to extend excursion set theory
by treating the critical threshold
for collapse  as a stochastic variable,  which
better captures some of the dynamical complexity of the halo formation phenomenon,  
 while in paper III
we  use the formalism developed in the present paper to compute 
the effect of non-Gaussianities
on the halo mass function. 

\end{abstract}



\section{Introduction}

The computation of the mass function of dark matter halos is a central
problem in modern cosmology. In particular, the high-mass tail of the
distribution is a  sensitive probe of primordial non-Gaussianities
\citep{MLB,MMLM,KOYAMA,MVJ,RB,RGS}. Various planned large-scale galaxy surveys,
both ground based (DES, PanSTARRS and LSST) and on satellite
(EUCLID and ADEPT) can detect the
effect of primordial non-Gaussianities on the mass distribution of
dark matter halos (see e.g. \cite{Dalal,CVM}). Of course, 
this also requires reliable theoretical predictions for the mass function,
first of all when the primordial fluctuations 
are taken to be gaussian, and then including 
non-Gaussian corrections.
Furthermore, the halo mass function  is both a
sensitive probe of cosmological parameters and a crucial ingredient 
when one studies the dark matter distribution, as well as
the formation, evolution
and distribution of galaxies, so its accurate  prediction 
is obviously important.

The formation and evolution of dark matter halos is a highly complex
dynamical process, and a detailed understanding of it can only come
through large-scale $N$-body simulations. 
Some analytical understanding is however also  desirable, both for
obtaining a
better physical intuition, and for the flexibility under changes of models or
parameters (such as cosmological model, shape of the non-Gaussianities,
etc.) that is the advantage of analytical results over very timing
consuming numerical simulations.

Analytic techniques generally
start by modelling the collapse as spherical or ellipsoidal. However,
$N$-body simulations show that the actual process of halo formation is 
not ellipsoidal, and in fact is not even a collapse, but rather a
messy mixture of violent encounters, smooth accretion and
fragmentation \citep{Springel:2005nw}. 
In spite of this,  analytical techniques 
based on Press-Schecther (PS) 
theory \citep{PS} and its extension known as excursion set
theory \citep{PH90,Bond} are able to reproduce, at least qualitatively, several properties of dark matter halos such as their conditional and unconditional
mass function,  halo accretion
histories, merger rates and halo bias (see 
\cite{Zentner} for a recent review). 
However, at the quantitative level,
already for gaussian fluctuations the prediction of
excursion set theory  for the mass function 
deviate significantly from the results of $N$-body simulations.
The halo mass function $dn/dM$
can be written as~\citep{jenkins2001}
\be\label{dndMdef}
\frac{dn(M)}{dM} = f(\s) \frac{\bar{\rho}}{M^2} 
\frac{d\ln\s^{-1} (M)}{d\ln M}\, ,
\ee
where  $n(M)$ is the number density of dark matter halos of mass $M$,
$\s^2$ is the variance of the linear density field smoothed on a
scale $R$ corresponding to a mass $M$, and
$\bar{\rho}$ is the average density of the universe. In 
excursion set theory within a spherical collapse model the function
$f(\s)$ is predicted to be

\begin{figure}
\includegraphics[width=0.45\textwidth]{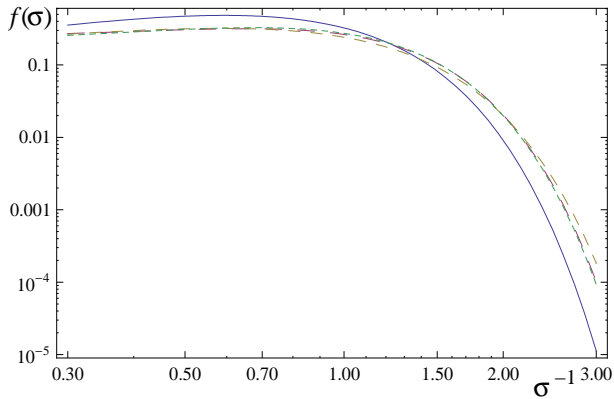}
\caption{\label{fig:fsigma_vari_fit}
A log-log plot of the function $f(\s)$. 
The (blue) solid curve is the PS prediction
$f_{\rm PS}(\s)$. The three almost
indistinguishable dashed lines are the  Sheth-Tormen fit to the
GIF simulation of \cite{kauf99}, and 
the  fit to
the $N$-body simulations of \cite{PPH} and \cite{Warren:2005ey}. 
The fitting functions are summarized in
Table~3 of \cite{PPH}.
}
\end{figure}

\noindent
\be\label{fps}
f_{\rm PS}(\s) = 
\(\frac{2}{\pi}\)^{1/2}\, 
\frac{\d_c}{\s}\, 
\, e^{-\d_c^2/(2\s^2)}\, ,
\ee
where $\d_c\simeq 1.686$ is the critical value for collapse
in the spherical collapse
model. This result can be extended to arbitrary redshift
$z$ reabsorbing the evolution of the variance into $\d_c$, so that
$\d_c$ in the above result is replaced by $\d_c(z)=\d_c(0)/D(z)$, where
$D(z)$ is the linear growth factor.
This prediction can be compared with the existing $N$-body
simulations 
(see e.g. \cite{jenkins2001,Warren:2005ey,Lukic:2007fc,
Tinker:2008ff,PPH,RKTZ} and
references therein).
The results of these simulations have been  represented by various fitting
functions, see e.g.
\cite{ST}, \cite{SMT}.
In Fig.~\ref{fig:fsigma_vari_fit} we compare the function 
$f_{\rm PS}(\s)$ given in \eq{fps}, 
to various  fits to $N$-body simulations, 
plotting the result
against $\s^{-1}$. High masses correspond to large
smoothing radius $R$, i.e.
low values of $\s$ and large $\s^{-1}$, so mass increases from left
to right on the horizontal axis. One sees that the $N$-body
simulations are quite consistent among them, and that
PS theory predicts too many low-mass halos, roughly by a factor of
two,  and too few high-mass halos: at $\s^{-1}=3$, PS theory is already
off by  a factor ${\cal O}(10)$.
The  primordial non-Gaussianities
can be constrained by probing the statistics of rare events, such as 
the formation of the most
massive objects,
so it is particularly important to
model  accurately the high-mass part of the halo mass function, 
first of all at the gaussian level.
It makes little sense to develop an analytic  theory of the
non-Gaussianities, by perturbing over a gaussian theory that in the
interesting mass range is already off by  one order of magnitude.

When searching for the origin of this failure of excursion set
theory, one can divide the possible concerns into two classes: 

(i) Even if one accepts 
as a physical model for halo formation a spherical (or ellipsoidal) collapse model,  there are
formal mathematical problems in the implementation of excursion set
theory that leads to \eq{fps}. 

(ii)  The physical model itself is inadequate, since
a spherical or even elliptical collapse model is an oversimplification of the actual complex process of halo formation.

Concerning point (i), it is well known that the original
argument of Press and Schechter  miscounts the number of virialized
objects because of the so-called
``cloud-in-cloud'' problem. In the spherical collapse model 
one assumes that  a region of radius
$R$, with a smoothed  density contrast $\d(R)$, collapses
and virializes once $\d(R)$ exceeds a critical value $\d_c\simeq
1.686$.\footnote{More precisely,  $\d_c$ has a slight dependence
on the cosmological model, and $\d_c=1.686$ is the value for a
$\Omega_M=1$ cosmology~\citep{Lacey:1993iv}. 
For a model with $\Omega_{M}+\Omega_{\Lambda}=1$
this dependence is computed in \cite{ECF}. For $\Omega_M\simeq 0.3$,
$\d_c$ is between $1.67$ and $1.68$, see their Fig.~1. This difference
is however much smaller than other uncertainties in our computation.}
Within PS theory, for gaussian
fluctuations the distribution
probability for the density contrast is
\be
\Pi_{\rm PS}(\d,S)=\frac{1}{\sqrt{2\pi S}}\, e^{-\d^2/(2S)}\, ,
\ee
where 
\be\label{defdiS}
S(R) \equiv \s^2(R)=
\langle \d^2(x,R)\rangle\, ,
\ee
and the fractional volume of space
occupied by virialized objects larger than $R$ is identified with
\be\label{F(M)PS}
F_{\rm PS}(R)=\int_{\d_c}^{\infty}d\d\, \Pi_{\rm PS}(\d, S(R))=
\frac{1}{2}\, {\rm erfc}\(\frac{\nu(R)}{\sqrt{2}}\)
\, ,
\ee
where $\nu(R) =\d_c/\s(R)$. 
As remarked already by Press and Schechter, this expression cannot however be
fully correct. In fact, in the hierarchical models that we are considering
the variance $S(R)$ diverges as $R\ra 0$, so all  the mass in the
universe must finally be contained in virialized objects.
Thus, we
should have $F_{\rm PS}(0)=1$, while \eq{F(M)PS} gives 
$F_{\rm PS}(0)=1/2$. Press and
Schechter corrected this simply adding by hand an overall factor of
two. 

The reason for this failure
is that the above procedure misses the cases in which, on a given
smoothing
scale $R$,  $\d(R)$ is below the threshold, but still it
happened to be above the threshold at some scale  $R'>R$. Such a configuration
corresponds to a virialized object of mass
$M'>M$. However, it is not counted in $F_{\rm PS}(R)$ 
since on the scale $S$ it is below threshold.
Thus \eq{F(M)PS}  cannot  be fully correct. 

In \cite{Bond}
this problem was solved by mapping the evolution of $\d$ with the smoothing
scale into a stochastic problem. Using a sharp $k$-space filter, they
were able to formulate the problem in terms of a Langevin equation
with a Dirac-delta noise. In other words, the smoothed 
density perturbation $\d$ suffers a markovian
stochastic motion under the influence of a  gaussian white
noise, with the 
variance $S=\s^2$ playing the role of a time variable.
In this formulation, the halo is defined to be formed when the
smoothed density
perturbation $\d$ reaches the critical value $\d_c$ for the first
time. The problem is therefore reduced to a
``first-passage problem'', which is a classical subject in the
theory of stochastic processes \citep{redner2001}. 
One may write a Fokker-Planck
equation describing the probability $\Pi(\d,S)$ that the density 
perturbation acquires a given value $\d$ at a given ``time'' $S$,
supplemented 
by the absorbing barrier boundary condition that the probability vanishes
when $\d=\d_c$. The  solution reproduces \eq{fps}, including the
factor of two  
that Press and 
Schechter were forced to introduce by hand.\footnote{The work of \cite{Epstein}
also solves the cloud in cloud problem and recovers the correct factor of two, though the process considered therein uses Poisson seeds for structure formation.}

However, this procedure still raises some technical questions, that will be
reviewed in more detail in Section~\ref{sect:Compu}. In short,  there
are two issues that  deserve a deeper scrutiny.
First, the   ``absorbing barrier'' boundary
condition $\Pi (\d_c,S)=0$  is 
a natural one, but still it is something that is imposed by hand, and
in this sense it is really an ansatz.
In the literature for stochastic 
processes it is  well-known that, in general, the probability does
not satisfy any simple boundary condition \citep{vKampen,knessl}.
This is due to the
fact that, when one works with a discretized time step, a  stochastic trajectory can exit a given domain by jumping 
over the boundary without hitting it, 
unlike a continuous diffusion process which has to hit
the boundary to exit the domain. Particular care must therefore be 
devoted to the passage from the discrete to the continuum. As we will see,
the passage from a discrete to a continuum formulation
is indeed highly non-trivial when a generic filter and/or non-Gaussian
perturbations are used.

A second related concern is
that the derivation of Bond et al. only works for 
a sharp $k$-space filter.
However, as we  review in Section~\ref{sect:Compu},
there is no unambiguous way of associating a mass to a region of size
$R$ smoothed with a sharp $k$-space filter. The only unambiguous way
of associating a mass $M$ to a smoothing scale $R$ is using  a sharp filter
in $x$-space, proportional to $\th(R-r)$, 
in which case one has the obvious relation 
$M=(4/3)\pi R^3\rho$.
This is also the relation used in numerical simulations. 
As soon as one uses a different  filter (such as the
tophat in real space), 
the Langevin equation with gaussian Dirac-delta noise, that describes a
simple markovian process, is replaced by a very complicated
non-markovian dynamics dictated by a colored noise. The system 
acquires memory properties and 
the probability $\Pi(\d,S)$ no longer satisfy a simple diffusion
equation such as the 
Fokker-Planck equation. The same
is true if the density perturbation is non-Gaussian. 
Furthermore,   the correctness of the 
``absorbing barrier'' boundary condition 
is now far from obvious.
These difficulties are well-known in the
statistical physics  community, where progress in solving the 
first-passage  problem in the presence of a non-markovian 
dynamics  has been very limited \citep{hanggi1981,weiss1983,vkampen1998}. 
From these considerations one concludes that
the rather common procedure of taking  the analytical results  
of \cite{Bond}, valid for a sharp filter in momentum space,
and applying them to    
generic filters is incorrect.\footnote{Similarly, even if the mathematical
problem of solving the Fokker-Planck equation with a moving barrier 
is amenable to an elegant formulation \citep{hui2006}, its application
to the halo mass function suffers from the problem that for a general filter 
it is incorrect to
assume  that the probability
$\Pi(\d,S)$ evolves according to the  Fokker-Planck equation.} 

These issues become even more important
when one considers the evolution with smoothing scale of non-gaussian
fluctuations, since non-gaussianities induce again a non-markovian
dynamics, and furthermore it is important to disentangle
the physically interesting non-markovian 
contribution to the halo mass function due to 
primordial non-gaussianities, from the  non-markovian
contribution due to the filter function.

Concerning point (ii) above, it is important to stress  once again that  excursion set theory
is just a simple mathematical model for a complex dynamical process. 
Treating the collapse as ellipsoidal rather than spherical  gives a more realistic description \citep{ST,SMT}. However,
as we already mentioned,  dark matter halos grow through a mixture of  smooth accretion,
violent encounters and fragmentations, and modeling halo collapse as spherical, or even ellipsoidal, is certainly an oversimplification. In addition, the very definition of 
what is a dark matter halo, both in $N$-body simulations and
observationally, is a difficult problem (for cluster observations, see
\cite{Jeltema} and references therein), that we will discuss
in more detail in paper~II.

In this series of paper we  examine systematically the above
issues. In the present paper we start from excursion set theory in its simpler 
physical implementation, i.e. 
coupled to a spherical collapse model, and within this framework we put the
formalism on firmer mathematical grounds. We show how to
formulate the mathematical problem exactly in terms of a path integral with
boundaries and particular care will be devoted to 
the passage from the discrete to the continuum. 
This formalism allows us to obtain a number of result:
first, when we restrict to gaussian fluctuations and sharp $k$-space filter, in the continuum limit we recover
the usual formulation of excursion set theory, but in this case the absorbing barrier boundary condition emerges
automatically from the
formalism, without the need of  imposing it by hand.
For different filters the problem becomes much more
complicated, and we have to deal with a non-markovian
dynamics. 
We will see that, for a generic filter, 
the zeroth-order term in an expansion of the
non-markovian contributions gives back
\eq{fps}, where $\s^2$ is now the variance computed with the generic  filter. 
We then
show how the non-markovian contributions 
can be computed perturbatively using our path
integral formulation,  and we
compute explicitly, to first perturbative
order, the halo mass function for
a tophat filter in coordinate space. We  find that the non-markovian 
contributions do not alleviate the discrepancy with $N$-body simulations. On
the contrary, in the relevant mass range
the full halo mass function is everywhere slightly 
lower than the one obtained from the markovian contribution, 
so in the large mass regime this correction goes in
the wrong direction. This result will not be a surprise to the expert
reader. Already in their classical paper, Bond et al.
computed the result with a tophat filter in coordinate space using a
Monte Carlo (MC) realization of the trajectories obtained from a
Langevin equation with colored noise, and 
found indeed that one has fewer high mass
objects. More recently, a MC simulation of this kind has been done in 
\cite{RKTZ}, and  our analytical result to first order
is in  agreement  with their findings.

In paper~II of this series,
motivated by  the physical limitations of the spherical or ellipsoidal collapse model,  we propose that some
of the physical complications of the realistic process of halo formation and growth can be included in the excursion set framework, at least at an effective level, by assuming that the critical value for collapse is not a fixed constant 
$\d_c$, as in the spherical collapse model, nor a fixed function of the variance $\s^2$, as in the ellipsoidal collapse model, but rather is itself a stochastic variable, whose scattering reflects a number of complicated aspects 
of the underlying dynamics.

Finally, in paper~III of this series we apply the formalism developed
in the present paper, together with the diffusing barrier model
developed in paper~II, 
to the computation of the halo mass function in the presence of
non-Gaussian fluctuations.

This paper is organized as follows. In 
Section~\ref{sect:Compu} we review the
excursion set theory developed in  \cite{Bond}; in 
Section~\ref{sect:micro} we present
the path integral approach to a stochastic problem in the presence of 
a barrier. In Sections~\ref{sect:Bond} we
specialize to the cases of a sharp filter in momentum space, while in
Section~\ref{sect:extensiongau} we consider
a generic filter. In particular, in Section~\ref{sect:extensiongau}
we show how to deal with the non-markovian corrections to the 
halo mass function. Some technicalities regarding the delicate
passage from the 
discrete to the continuum 
are contained in Appendices A and B. 

\section{The computation of the halo mass function as a stochastic problem}
\label{sect:Compu}

The computation of the halo mass
function can be formulated in terms of a
stochastic process, as is well known since the classical 
work of \cite{Bond}. Let us recall the procedure, in order to set the
notation and to highlight some delicate points, in particular related to the
choice of the filter function, that are important in the
following. The expert reader might wish to move directly to 
Section~\ref{sect:micro}.

\begin{figure}
\includegraphics[width=0.4\textwidth]{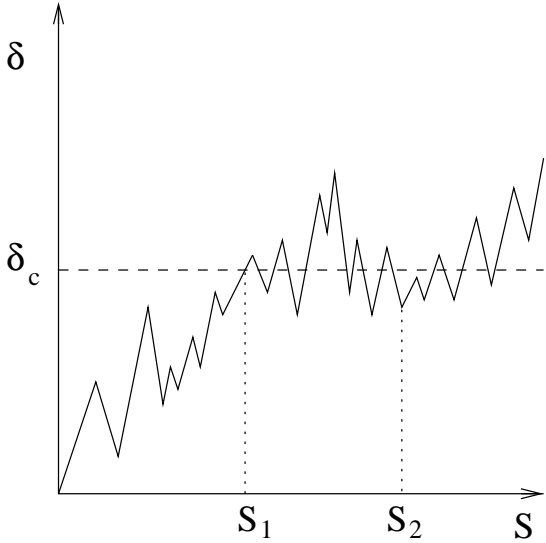}
\caption{\label{fig:traj}
A trajectory that performs multiple up-crossings of the threshold at
$\d=\d_c$. 
}
\end{figure}

One considers 
the density contrast $\d({\bf x}) = [\rho ({\bf x})-\bar{\rho}]/\bar{\rho}$,
where $\bar{\rho}$ is the mean mass density of the universe and ${\bf x}$
is the comoving position,  and  smooths it
on some scale $R$, defining
\be\label{dfilter}
\d({\bf x},R) =\int d^3x'\,  W(|{\bf x}-{\bf x}'|,R)\, \d({\bf x}')\, ,
\ee
with a filter function 
$W(|{\bf x}-{\bf x}'|,R)$. We denote by
$\tilde{W}({\bf k},R)$ its Fourier transform.
A simple  choice is a sharp filter in $k$-space,
\be\label{Wsharpk}
\tilde{W}_{{\rm sharp}-k}(k,k_f)=\theta(k_f-k)\, ,
\ee
where $k_f=1/R$, $k=|{\bf k}|$ and $\theta$ is the step function. Other
common choices are
a  sharp filter in $x$-space,
$\tilde{W}_{{\rm sharp}-x}(r,R)= [3/(4\pi R^3)]\theta(R-r)$,
or a gaussian filter,
$\tilde{W}_{\rm gau}(k,R)= e^{-R^2k^2/2}$.
Writing  \eq{dfilter} in terms  of the Fourier transform we have
\be\label{dxdkWk}
\d({\bf x},R) =\int \frac{d^3k}{(2\pi)^3}\,
\tilde{\d}({\bf k} ) \tilde{W}(k,R) e^{-i{\bf k\cdot x}}\, ,
\ee
where $k=|{\bf k}|$.
We focus on the evolution of $\d({\bf x},R)$ with $R$ at a fixed
value of ${\bf x}$, that we can choose without loss of generality as 
${\bf x}=0$, and we write $\d({\bf x}=0,R)$ simply as 
$\d (R)$.
Taking the derivative of \eq{dxdkWk} with respect to $R$ we get
\be\label{padeta1}
\frac{\pa\d(R)}{\pa R} =\zeta (R)\, ,
\ee
where
\be\label{zetadelta}
\zeta(R)\equiv \int \frac{d^3k}{(2\pi)^3}\,
\tilde{\d}({\bf k} ) \frac{\pa\tilde{W}(k,R)}{\pa R}\, .
\ee
Since the modes $\tilde{\d}({\bf k} )$ are stochastic variables,
$\zeta(R)$ is a stochastic variable too, and
\eq{padeta1} has the form of a Langevin
equation, with $R$ playing the role of time, and
$\zeta(R)$ playing the role of noise. 
When $\d(R)$ is a gaussian
variable,  only its two-point connected correlator is
non-vanishing. In this case, we see from  \eq{zetadelta} that
also $\zeta$ is gaussian.  The two-point function of $\d$ defines 
the power spectrum $P(k)$,
\be\label{defP(k)}
\langle \tilde{\d}({\bf k} )\tilde{\d}({\bf k}')\rangle =
(2\pi)^3\d_D({\bf k}+{\bf k}') P(k)\, .
\ee
From this it follows that
\be\label{corr2eta}
\langle \zeta(R_1)\zeta(R_2)\rangle =
\int_{-\infty}^{\infty}d(\ln k)\,\, \D^2(k)
\frac{\pa\tilde{W}(k,R_1)}{\pa R_1}
\frac{\pa\tilde{W}(k,R_2)}{\pa R_2}\, ,
\ee
where, as usual, $\D^2(k)=k^3P(k)/(2\pi^2)$. 
For a generic filter function the right-hand side is a 
function of  $R_1$ and $R_2$, different from a Dirac delta 
$\d_D(R_1-R_2)$. In the literature on stochastic processes this case 
is known as {\em  colored gaussian noise}. Things simplify
considerably for a sharp $k$-space filter. Using $k_f=1/R$ instead of
$R$, and defining $Q (k_F)=-(1/k_F) \zeta(k_F)$,
\eqs{padeta1}{corr2eta} become
\be
\label{padeta2}
\frac{\pa\d(k_F)}{\pa \ln k_F} =Q (k_F)\, ,
\ee
and
\be\label{corr2etab}
\langle Q ({k_F}_1)Q ({k_F}_2)\rangle =
\D^2({k_F}_1) \d_D(\ln {k_F}_1-\ln {k_F}_2)\, .
\ee
Therefore, we have a Dirac delta noise.
We can write these equations in an even simpler form using as ``pseudotime''
variable the  variance $S$ defined in \eq{defdiS}. Using \eq{dxdkWk}
\be\label{mu2RW2}
S(R) =
\int_{-\infty}^{\infty} d(\ln k)\, \D^2(k) 
|\tilde{W}(k,R)|^2\, .
\ee
For a sharp $k$-space filter, $S$ becomes
\be
S(k_F) =\int_{-\infty}^{\ln k_F} d(\ln k)\, \D^2(k) \, ,
\ee
so
\be
\frac{\pa S}{\pa\ln k_f} =\D^2(k_f)\, .
\ee
Thus, redefining finally $\eta (k_F) =Q(k_F)/\D^2(k_F)$, we get
\be\label{Langevin1}
\frac{\pa\d(S)}{\pa S} = \eta(S)\, ,
\ee
with
\be\label{Langevin2}
\langle \eta(S_1)\eta(S_2)\rangle =\d (S_1-S_2)\, .
\ee
which is a the  Langevin equation with Dirac-delta noise, with 
$S$ playing the role of time. 
In hierarchical power spectra,
at $R=\infty$ we have $S=0$,
and $S$ increases monotonically
as $R$ decreases. Therefore we can start from $R=\infty$,
corresponding to ``time'' $S=0$, where $\d=0$, and  follow the
evolution of $\d(S)$ as we decrease $R$, i.e. as we increase
$S$.
 The fact that this evolution is governed by the Langevin equation
means that  
$\d(S)$ performs  a random
walk, with respect to the ``time'' variable $S$. 
Following \cite{Bond}, we 
refer to the evolution of $\d$ as a function of $S$ as a ``trajectory''. 
In the spherical collapse model, a virialized object forms as soon as
the trajectory exceeds the threshold $\d=\d_c$. In this language, the 
``cloud-in-cloud''  problem of PS theory
is associated with trajectories that make multiple
crossings of the  threshold, such as that shown in Fig.~\ref{fig:traj}. 
If we compute the probability
distribution at $S=S_2$  as in PS theory, i.e. using \eq{F(M)PS},
this trajectory does not contribute to $F_{\rm PS}(R)$
since at
this value of $S$ it is below threshold. However, it has already gone
above threshold at an earlier time $S_1$, corresponding to a radius
$R_1$, so it gives a
virialized object of mass $M(R_1)>M(R_2)$. This virialized object has
been lost in $F_{\rm PS}(R_2)$ evaluated through
\eq{F(M)PS}, in spite of the fact that this formula
was supposed to count all objects with mass greater then $M(R_2)$.

To cure the  ``cloud-in-cloud'' problem  
we must consider  the {\em lowest }
value of $S$ (or, equivalently, the highest value
of $R$) for which the trajectory pierces the threshold. Similar
problems are known in statistical physics as ``first-passage time'' 
problems.
After that, a virialized object forms and this trajectory should be
excluded from further consideration. 
We therefore consider an ensemble 
of trajectories, all starting from the
initial value $\d=0$ at initial ``time'' $S=0$,  and 
we compute the function  $\Pi (\d,S)$
that gives the probability distribution of
reaching a value $\d$ at
``time'' $S$. As is well known, 
if a stochastic process obeys the Langevin equation
(\ref{Langevin1}) with a Dirac delta noise (\ref{Langevin2}), the
corresponding distribution function is a solution of the Fokker-Planck
(FP) equation,
\be\label{FPdS}
\frac{\pa\Pi}{\pa S}=\frac{1}{2}\, \frac{\pa^2\Pi}{\pa \d^2}\, .
\ee
We denote by $\Pi^0 (\d,S)$
the solution of this equation over the whole real axis 
$-\infty<\d<\infty$, with the boundary condition that it vanishes at
$\d=\pm\infty$. One can check immediately that
\be\label{singlegau}
\Pi^0 (\d,S)=\frac{1}{\sqrt{2\pi S}}\, e^{-\d^2/(2S)}\, .
\ee
This probability distribution
would bring us back to PS theory, and to its problems discussed
in the Introduction. So,
we  need to eliminate the trajectories once they have
reached the threshold. In \cite{Bond} this is implemented by
imposing the boundary condition
\be
\left.\Pi (\d,S)\right|_{\d=\d_c}=0\, .
\ee
This seems very natural, but we stress that this boundary condition is
still something that it is imposed by hand. 
The solution of the FP equation with this boundary condition
is~\citep{Chandra} 
\be\label{PiChandra}
\Pi (\d,S)=\frac{1}{\sqrt{2\pi S}}\,
\[  e^{-\d^2/(2S)}- e^{-(2\d_c-\d)^2/(2S)} \]
\, ,
\ee
and gives the distribution function of  excursion set theory.
When studying halo merger trees it is important to consider also the
distribution for trajectories that start from an arbitrary value
$\d_0\neq 0$~\citep{Bond,Lacey:1993iv}. In this case, \eq{PiChandra} is
replaced by
\be\label{PiChandra2}
\Pi (\d_0;\d;S)=\frac{1}{\sqrt{2\pi S}}\,
\[  e^{-(\d-\d_0)^2/(2S)}- e^{-(2\d_c-\d_0-\d)^2/(2S)} \]
\, .
\ee
This result is easily
understood  writing
$2\d_c-\d_0-\d = 2(\d_c-\d_0) - (\d-\d_0)$,
so \eq{PiChandra2} is obtained from \eq{PiChandra} performing the
obvious replacement 
$\d\ra \d-\d_0$,  and also $\d_c\ra \d_c-\d_0$, which  expresses
the fact that, if we start from $\d_0$, the random walk must cover a
distance $\d_c-\d_0$ to reach the threshold.

In the excursion set theory the distribution $\Pi (\d,S)$ 
is defined only for $\d<\d_c$, so 
the fraction $F(S)$ of trajectories that have crossed the threshold at
``time'' smaller or equal to $S$ cannot be written, as in
\eq{F(M)PS}, as an integral from $\d=\d_c$ to $\d=+\infty$. 
Rather, we use the fact that the integral of
$\Pi (\d,S)$  from $\d=-\infty$ to $\d=\d_c$ gives the fraction of
trajectories that at ``time'' $S$ have never crossed the threshold, so
\be\label{F(S)1menoint}
F(S)=1-\int_{-\infty}^{\d_c}d\d\, \Pi (\d,S)\, .
\ee
Observing that $\Pi (\d,S)=\Pi^0 (\d,S) - \Pi^0 (2\d_c-\d,S)$,  we see
that
\be\label{FS1meno}
F(S)= 1-\int_{-\infty}^{\d_c}d\d\, \Pi^0 (\d,S) 
+\int_{-\infty}^{\d_c}d\d\, \Pi^0 (2\d_c-\d,S)\, .
\ee
Since $\Pi^0 (\d,S)$ is normalized to one,
\be
1-\int_{-\infty}^{\d_c}d\d\, \Pi^0 (\d,S) =\int_{\d_c}^{\infty}d\d\,
\Pi^0 (\d,S)\, .
\ee
For the last term in \eq{FS1meno}, we
write $\d'=2\d_c-\d$, and
\be
\int_{-\infty}^{\d_c}d\d\, \Pi^0 (2\d_c-\d,S)=
\int_{\d_c}^{\infty}d\d'\,
\Pi^0 (\d',S)\, .
\ee
Thus, one obtains
\be\label{FSexcu}
F(S)=2\int_{\d_c}^{\infty}d\d\,\Pi^0 (\d,S)
=  {\rm erfc}\(\frac{\nu}{\sqrt{2}}\)
\, ,
\ee
where $\nu =\d_c/\s(M)$,
and one recovers the factor of two that Press and Schechter were
forced to introduce by hand. The probability of first
crossing the threshold between ``time'' $S$ and $S+dS$ is given by
${\cal F}(S) dS$, with
\be\label{defcalF}\label{firstcrossT}
{\cal F}(S) \equiv
\frac{dF}{dS} = -\int_{-\infty}^{\d_c}d\d\, \frac{\pa\Pi}{\pa S}\, .
\ee
This can be easily computed by making use of the fact that
$\Pi$ by definition satisfies the FP equation (\ref{FPdS}), so
\be\label{calF}
{\cal F}(S)=-\frac{1}{2} \, \left.\frac{\pa\Pi}{\pa\d}\right|_{\d=\d_c}
=\frac{\d_c}{\sqrt{2\pi}\, S^{3/2}}\, e^{-\d_c^2/(2S)}
\, .
\ee
Observe that, in $\d=\d_c$, $\Pi (\d,S)$ and all its derivative of
even order with respect to $\d$ vanish, while all its derivative of
odd order with respect to $\d$ are twice as large as the 
value for the single
gaussian (\ref{singlegau}). So, this first-crossing rate is twice as
large as that computed with a single gaussian, which is another why of
understanding how one gets the factor of two that the original form
of PS theory misses.

The halo mass function follows if one has a relation $M=M(R)$ that gives
the mass associated to the smoothing of $\d$ over a region of
radius $R$. We  discuss below
the subtleties associated to this relation, and its dependence on the
filter function. Anyhow, once $M(R)$ is
given, we can consider $F$ as a function of $M$ rather than of
$S(R)$. Then $|dF/dM| dM$ is the fraction of volume occupied by
virialized objects with mass  between $M$
and $M+dM$. Since each one occupies a volume $V=M/\bar{\rho}$, where
$\bar{\rho}$ is the average density of the universe, the
number of virialized object $n(M)$ with mass between $M$
and $M+dM$ is given by
\be
\frac{dn}{dM} dM=\frac{\bar{\rho}}{M}\, \left|\frac{dF}{dM}\right| dM\, ,
\ee
so
\be
\frac{dn}{dM}=\frac{\bar{\rho}}{M}\,\frac{dF}{dS}
\left|\frac{dS}{dM}\right| =
\frac{\bar{\rho}}{M^2}{\cal F}(S) 2\s^2\frac{d\ln\s^{-1}}{d\ln M}\, ,
\ee
where we used $S=\s^2$.
Therefore,
in terms of the first-crossing rate ${\cal F}(S)=dF/dS$, the function
$f(\s)$ defined from \eq{dndMdef} is given by
\be\label{fcalF}
f(\s)= 2\s^2{\cal F}(\s^2)\, .
\ee
Using \eq{calF} we get the halo mass function
in  PS theory (with the factor of two computed thanks to the excursion
set theory),
\be
\(\frac{dn}{dM}\)_{\rm PS}
=\(\frac{2}{\pi}\)^{1/2}\,\frac{\d_c}{\s}\, e^{-\d_c^2/(2\s^2)}
\, \frac{\bar{\rho}}{M^2} \frac{d\ln\s^{-1}}{d\ln M}
\, ,
\ee
This is the result given in
\eqs{dndMdef}{fps}.

The crucial point is how to associate  a mass $M$ to the 
filter scale $R$. For the  sharp filter in $x$-space this is 
clear. The mass associated to a
spherical region of radius $R$ and density  $\rho$
is $M=(4/3)\pi R^3\rho$. For the other filters there is no 
unambiguous definition. A possibility often used is the following. One first
normalizes $W$ so that its maximum value is one. Calling
$W'$ this new dimensionless filter, one can define the volume $V$
associated to the filter as
$V=\int d^3x\, W'$,
and $M=\rho V$. This procedure  seems reasonable,
but still it is somewhat arbitrary, since one might as well chose a
different normalization for $W'$. For a gaussian filter, this gives 
$V=(2\pi)^{3/2}R^3$. For a sharp $k$-space filter, on top of this
ambiguity, there is also the fact that such a volume is not even well
defined. In fact, the $k$-space filter
in coordinate space reads
\be\label{Wsharpkxspace}
W_{{\rm sharp}-k}(r,R) = \frac{1}{2\pi^2 R^3}\, 
\frac{\sin u - u\cos u}{u^3}\, ,
\ee
where $u=r/R=k_f r$ and $r=|{\bf x}-{\bf x}'|$, which gives
\be
W' = 3\, \frac{\sin u - u\cos u}{u^3}\, ,
\ee
and
\be
V = 4\pi\int_0^{\infty}dr\, r^2 W' 
=12\pi R^3\int_0^{\infty}du\,\[ \frac{\sin u}{u} -\cos u\]\, .
\ee
The integral of $\sin u/u$ gives $\pi/2$, but the limit for 
$\Lambda\ra \infty$ of
the integral of $\cos u$ from $u=0$ to $u=\Lambda$ does not exist.
If one just sets it to zero, without much justification, one finds the result
$V=6\pi^2R^3$ which is sometimes quoted~\citep{Lacey:1993iv}. In any case, it is
clear that it is difficult to give unambiguous numerical predictions
for the halo mass function with a filter different from the sharp
$x$-space filter.

The standard practice  in the literature is to use the
PS mass function,
which can  be derived 
from excursion set theory but only if one works
with a sharp $k$-space filter, and at the same time to use
$M=(4/3)\pi R^3\rho$, which is only
valid for a sharp $x$-space filter. Of course this is not
consistent and cannot be a good starting point for the inclusion
of  the non-Gaussianities, since one would
attribute to primordial 
non-Gaussianities features in the mass function which are
due, more trivially, to the filter function.

In principle, one can determine the halo mass function with a 
tophat filter in coordinate space by performing a Monte Carlo
(MC) realization of the trajectories obtained from a
Langevin equation with colored noise \citep{Bond,RKTZ}.  
However, our final aim is to get
some analytic understanding of the effect of non-Gaussianities on
the halo mass function, and
to this purpose we need 
a good analytic control of the effect of the filter, first of all in the
gaussian case.

\section{Path integral approach to stochastic problems}\label{sect:micro}

\subsection{General formalism}

We have seen that the computation of the halo mass
function can be reformulated in terms
of a stochastic process. 
We now show how to compute the probability distribution of a
variable evolving stochastically, in terms of 
its correlators. In this paper we  limit ourselves to gaussian
variables, while in paper~III of this series we  perform the
generalization to arbitrary non-Gaussian theories.

Let us  consider a variable
$\d(S)$  that evolves stochastically with ``time'' $S$,
with zero mean $\langle\d(S)\rangle =0$.
For a gaussian theory, the only
non-vanishing connected correlator is then the two-point correlator
$\langle \delta (S_1)\delta (S_{2})\rangle_c$, 
where the subscript $c$ stands for
connected.  

We consider an ensemble of
trajectories all starting at $S_0=0$ from an initial position
$\d(0) =\delta_0$, 
and we follow them for a
time $S$.  
We discretize the interval $[0,S]$ in steps
$\D S=\eps$, so $S_k=k\eps$ with $k=1,\ldots n$, and $S_n\equiv S$. 
A trajectory is  defined by
the collection of values $\{\delta_1,\ldots ,\delta_n\}$, 
such that $\delta(S_k)=\delta_k$.
There is no absorbing barrier, i.e. $\delta(S)$ is allowed to range freely from
$-\infty$ to $+\infty$.
The probability density in the space of  trajectories is 
\be\label{defW}
W(\delta_0;\delta_1,\ldots ,\delta_n;S_n)\equiv \langle
\d_D (\delta(S_1)-\delta_1)\ldots \d_D (\delta(S_n)-\delta_n)\rangle\, ,
\ee
where, to avoid confusion with the density contrast $\d$, we
denote the Dirac delta by  $\d_D$.
In terms of $W$ we define
\be\label{defPi}
\Pi_{\eps} (\delta_0;\delta_n;S_n)
 \equiv\int_{-\infty}^{\delta_c} d\delta_1\ldots 
\int_{-\infty}^{\delta_c}d\delta_{n-1}\, 
W(\delta_0;\delta_1,\ldots ,\delta_{n-1},\delta_n;S_n)\, ,
\ee
where  $S_n=n\eps$.
So, 
$\Pi_{\eps} (\delta_0;\d;S)$ is the probability density of arriving at the
"position" $\d$ in a "time" $S$,   starting from $\delta_0$ at time $S_0=0$, 
through trajectories that never exceeded $\delta_c$.
Observe that the final point
$\d$  ranges over $-\infty<\d<\infty$. For later use, we
find useful to write explicitly 
that  $\Pi$ depends also on the temporal
discretization step $\eps$.
We are finally interested in its continuum limit, $\Pi_{\eps=0}$, and
we will see in due course that taking the limit $\eps\ra 0$ 
of $\Pi_{\eps}$ is non-trivial.

The usefulness of $\Pi_{\eps}$ is that it allows us to compute the
first-crossing rate from first principles, without the need of
postulating the existence of an absorbing barrier. Simply,
the quantity
\be
\int_{-\infty}^{\delta_c}d\delta\, \Pi_{\eps}(\delta_0;\d;S)
\ee
gives the probability that at time $S$ a trajectory always stayed in
the region $\d<\delta_c$, for all times smaller than $S$. The rate of change
of this quantity is therefore equal to minus the rate at which trajectories
cross for the first time the barrier, so 
the first-crossing rate is
\be
{\cal F}(S)=-\int_{-\infty}^{\delta_c}d\delta\,\pa_S \Pi_{\eps}(\delta_0;\d;S)
\ee
(where $\pa_S=\pa/\pa S$), just as in \eq{defcalF}.
The halo mass function is then obtained from this first-crossing rate
using \eqs{dndMdef}{fcalF}.
Observe that no reference to a hypothetical ``absorbing barrier'' is made in
this formalism.  We will discuss below how, and under what
conditions,  an effective
absorbing barrier emerges from this microscopic approach.

To express  $\Pi_{\eps} (\delta_0;\d;S)$,
in terms of
the two-point correlator  of the theory we
use the integral representation of the Dirac delta
\be
\d_D (x)=\inT\frac{d\lambda}{2\pi}\, e^{-i\lambda x}\, ,
\ee
and we write \eq{defW} as
\be\label{compW}
W(\delta_0;\delta_1,\ldots ,\delta_n;S_n)=\inT\frac{d\lambda_1}{2\pi}\ldots\frac{d\lambda_n}{2\pi}\, 
e^{i\sum_{i=1}^n\lambda_i\delta_i} \langle e^{-i\sum_{i=1}^n\lambda_i\delta(S_i)}\rangle
\, .
\ee
Observe that 
the dependence on $\delta_0$ here is hidden in the correlators of $\d$, e.g.
$\langle\d^2(S=0)\rangle =\d^2_0$. It is convenient to set for simplicity
$\delta_0=0$ in the intermediate computations, and it will be easy to
restore it in the final results.
For gaussian fluctuations,
\be
\langle e^{-i\sum_{i=1}^n\lambda_i\delta(S_i)}\rangle=
e^{ -\frac{1}{2}\sum_{i,j=1}^n\lambda_i\lambda_j
\langle\delta(S_i)\delta(S_j)\rangle_c}\, ,
\ee
as can be checked immediately by performing the Taylor expansion of the
exponential on  the left-hand side, and using the fact that, for
gaussian fluctuations, the generic correlator factorizes into sum of
products of two-points correlators.
This gives
\be\label{WnNG}
W(\delta_0;\delta_1,\ldots ,\delta_n;S_n)=\Dl\, \,
e^{ i\sum_{i=1}^n\lambda_i\delta_i
-\frac{1}{2}\sum_{i,j=1}^n\lambda_i\lambda_j
\langle\delta_i\delta_j\rangle_c}
\, ,
\ee
where
\be
\int{\cal D}\lambda \equiv
\inT\frac{d\lambda_1}{2\pi}\ldots\frac{d\lambda_n}{2\pi}\, ,
\ee
and $\delta_i\equiv\delta(S_i)$. Then
\bees\label{Piexplicit}\label{3defPi0xiixij}
\Pi_{\eps}(\delta_0;\delta_n;S_n)&=&\int_{-\infty}^{\delta_c} d\delta_1\ldots d\delta_{n-1}\,
\Dl\nn\\
&&\exp\left\{ i\sum_{i=1}^n\lambda_i\delta_i
-\frac{1}{2}\sum_{i,j=1}^n\lambda_i\lambda_j
\langle\delta_i\delta_j\rangle_c\right\}
\, .
\ees

\subsection{Gaussian fluctuations with sharp $k$-space filter}
\label{subsect:gausharp}

As we have seen in Sect.~\ref{sect:Compu},
the computation of the halo
mass function in the excursion set formalism with sharp $k$-space
filter can be reduced to a Langevin equation with a Dirac-delta
noise. Therefore, we now study
the case in which $\d$ has
gaussian statistics (so only the two-point connected function is
non-vanishing) and obeys 
the Langevin equation (\ref{Langevin1})
with
a noise $\eta (S)$ whose correlator is a Dirac delta,
\eq{Langevin2}.
Using  as initial condition
$\delta_0=0$, \eq{Langevin1} integrates to
\be
\delta(S)=\int_0^S dS'\, \eta(S')\, , 
\ee
so the 2-point correlator
is given by
\bees
\langle\delta(S_i)\delta(S_j)\rangle_c&=&\int_0^{S_i}dS\int_0^{S_j}dS'
\langle\eta(S)\eta(S')\rangle\\
&=&{\rm min}(S_i,S_j)=\eps\min(i,j) \equiv \eps A_{ij}\label{twopt}\, .\nn
\ees
Denoting by $W^{\rm gm}$ the value of $W$ 
when $\d$ is a gaussian variable and
performs a markovian random walk with respect to the smoothing scale, i.e. satisfies
\eqs{Langevin1}{Langevin2}, we get
\bees
&&W^{\rm gm}(\delta_0;\delta_1,\ldots ,\delta_n;S_n)
=\inT\frac{d\lambda_1}{2\pi}\ldots\frac{d\lambda_n}{2\pi}\nn\\
&&\hspace{2cm}\times\exp\{i\sum_{i=1}^n\lambda_i\delta_i -\frac{\eps}{2} \sum_{i,j=1}^n
A_{ij}\lambda_i\lambda_j\}\nn\\
&&=
\frac{1}{(2\pi\eps)^{n/2}}\, \frac{1}{({\rm det}\, A)^{1/2}}
\exp\left\{-\frac{1}{2\eps}\, \sum_{i,j=1}^n\delta_i (A^{-1})_{ij}\delta_j\right\}\, .
\label{preW}
\ees
Given that $A_{ij}={\rm min}(i,j)$, we can verify that $A^{-1}$ is as follows: 
$(A^{-1})_{ii}=2$ for $i=1, \ldots, n-1$, $(A^{-1})_{nn}=1$, and
$(A^{-1})_{i,i+1}=(A^{-1})_{i+1,i}=-1$, for $i=1,\ldots, n-1$,
while all other matrix elements are zero. Furthermore,
${\rm det}\,\, A=1$. As a result, we
get
\bees\label{Wold}
&&\hspace*{-5mm}
W^{\rm gm}(\delta_0=0;\delta_1,\ldots ,\delta_n;S_n)=\frac{1}{(2\pi\eps)^{n/2}}\, \nn\\
&&\times\exp\left\{-\frac{1}{2\eps}\,\[
\delta_n^2+2\sum_{i=1}^{n-1} \delta_i (\delta_i-\delta_{i+1})\]\right\}\, .
\ees
This expression takes a more familiar form using the identity
$2 \delta_i (\delta_i-\delta_{i+1}) = (\delta_{i+1}-\delta_i)^2 - (\delta_{i+1}^2-\delta_i^2)$, together
with
$\sum_{i=1}^{n-1} (\delta_{i+1}^2-\delta_i^2) = \delta_n^2-\delta_1^2$.
Recall also that \eq{twopt} assumed as initial condition
$\delta_0=0$. The result for $\delta_0$ generic is simply obtained 
by replacing $\delta_i\ra \delta_i-\delta_0$ for all $i>0$. Then, 
for $i>0$ the terms $(\delta_{i+1}-\delta_i)^2$ are unaffected, while in the last
term of the sum $\delta_1^2\ra \delta_1^2-\delta_0^2$. Thus, for $\delta_0$ arbitrary,
we  get 
\be\label{W}
W^{\rm gm}(\delta_0;\delta_1,\ldots ,\delta_n;S_n)=\frac{1}{(2\pi\eps)^{n/2}}\, 
\exp\left\{-\frac{1}{2\eps}\,
\sum_{i=0}^{n-1}  (\delta_{i+1}-\delta_i)^2\right\}\, .
\ee
Observe that $W^{\rm gm}(\delta_0;\delta_1,\ldots ,\delta_n;S_n)d\delta_1\ldots d\delta_{n-1}$
is just the Wiener measure (see e.g. chapter~1 of  
\cite{Chaichian:2001cz}).
From \eq{W}
we see that
\be\label{markov}
W^{\rm gm}(\delta_0;\delta_1,\ldots ,\delta_n;S_n)=
\Psi_{\eps}(\delta_n-\delta_{n-1})
W^{\rm gm}(\delta_0;\delta_1,\ldots ,\delta_{n-1};S_{n-1})\, , 
\ee
where
\be\label{Psi}
\Psi_{\eps}(\D \d)=\frac{1}{(2\pi\eps)^{1/2}}\, 
\exp\left\{-\frac{(\D \d)^2}{2\eps}\, \right\}\, .
\ee
\Eq{markov} expresses the fact that the evolution determined by
\eqs{Langevin1}{Langevin2} is
a markovian process, i.e. the probability of jumping from the position
$\delta_{n-1}$ at time $S_{n-1}$ to the position
$\delta_{n}$ at time $S_{n}$  depends only on the values of
$\delta_n-\delta_{n-1}\equiv \D \d$ and on
$S_n-S_{n-1}\equiv \eps$, and not on the past history of the trajectory.
Integrating \eq{markov}
over $\delta_1, \ldots, \delta_{n-1}$ from  $-\infty$ to $\delta_c$ we get the
important relation
\be\label{CK2b}
\Pi^{\rm gm}_{\eps} (\delta_0;\delta_n;S_n)
=\int_{-\infty}^{\delta_c} d\delta_{n-1}\Psi_{\eps}(\delta_n-\delta_{n-1})
\Pi^{\rm gm}_{\eps} (\delta_0;\delta_{n-1};S_{n-1})\, ,
\ee
which generalizes the well-known 
Chapman-Kolmogorov equation to the case of  finite $\delta_c$.

\section{Derivation of  excursion set formalism for gaussian
  fluctuations and sharp $k$-space filter}
\label{sect:Bond}

We  now want to 
derive, from our ``microscopic''
approach,   the 
excursion set formalism of  \cite{Bond}.
As we have seen in Section~\ref{sect:Compu}, the result of
Bond et al. holds for 
gaussian fluctuations and sharp $k$-space filter, working
directly  in
the continuum limit, and reads
\be\label{Pix0}
\Pi^{\rm gm}_{\eps=0} (\delta_0;\d;S)
= \frac{1}{\sqrt{2\pi S}}\,
\[e^{-(\d-\delta_0)^2/(2S)}-
e^{-(2\delta_c-\delta_0-\d)^2/(2S)}\]\, .
\ee
We want to prove \eq{Pix0}
using our definition of $\Pi_{\eps}$ as a path
integral over all trajectories that never exceed $\delta_c$. 
Beside being a starting point for the generalization to arbitrary filter
functions and to non-Gaussian
theories, the derivation of the excursion set theory from first
principles has an intrinsic interest. In fact,  in
\cite{Bond} this
result is obtained by {\em postulating} that  the distribution function obeys a
FP equation with an ``absorbing barrier'' boundary condition
$\Pi(\delta_0;\d;S)|_{\d=\delta_c}=0$. While the fact that $\Pi_{\eps=0}$ obeys a
FP equation follows from \eq{Langevin1}, the 
absorbing barrier boundary condition 
is rather imposed by hand.  As we already mentioned, in the literature
on stochastic processes it is known that,
in the general case,  the 
distribution function $\Pi_{\eps} (\delta_0;\d;S)$ does not satisfy any
simple boundary condition  \citep{vKampen,knessl}.
It is therefore  
interesting to see how, in the gaussian case with sharp $k$-filter, 
an absorbing barrier
boundary condition effectively emerges from our microscopic approach.

We  first show that in the continuum limit we recover 
\eq{Pix0}. Then, we 
examine the finite-$\eps$ corrections. As it  turns out,
these corrections have a non-trivial structure which is quite
interesting in itself. Our main reason for discussing them in  detail,
however, is that they
play a crucial role in
the extension of our formalism to  a generic filter function and
to non-Gaussian fluctuations.

\subsection{The continuum limit}\label{subsect:cont}

To compute $\Pi^{\rm gm}_{\eps}$
by performing directly the integrals over 
$\delta_1, \ldots ,\delta_{n-1}$ in \eq{defPi},
and then taking the limit $\eps\ra 0$
is very difficult, 
since the integrals in \eq{defPi}
run only up to $\delta_c$, and already the inner
integral gives an error function whose argument
involves the next integration variable.

A better strategy is to make use of
\eq{CK2b}. This relation expresses the fact that,
for gaussian fluctuations and
sharp $k$-space filter, the underlying
stochastic process is markovian. 
We change notation, denoting $\delta_n=\d$, $\delta_n-\delta_{n-1}=\D \d$, and
$S_{n-1}=S$, so $S_n=S+\eps$. For fixed $\d$, we  have 
$d\delta_{n-1}=-d(\D\d)$, and \eq{CK2b} becomes\footnote{In this section we
  always assume that $\delta_0$ is strictly smaller than $\delta_c$. The case
  $\delta_0=\delta_c$ is important when we study the non-markovian
  corrections, and will be examined in due course.}
\be\label{CK3}
\Pi^{\rm gm}_{\eps} (\delta_0;\d;S+\eps)=
\int_{\d-\delta_c}^{\infty} d(\D\d)\, \Psi_{\eps}(\D\d)
\Pi^{\rm gm}_{\eps} (\delta_0;\d-\D \d;S)\, .
\ee
In the limit $\eps\ra 0$ we have
$\Psi_{\eps}(\D\d)\ra \d_D(\D\d)$, so to zeroth order in $\eps$
\eq{CK3} gives
\be
\Pi^{\rm gm}_{\eps=0} (\delta_0;\d;S)=
\int_{\d-\delta_c}^{\infty} d(\D\d)\, \d_D(\D\d)
\Pi^{\rm gm}_{\eps=0} (\delta_0;\d-\D\d;S)\, .
\ee
If $\d-\delta_c<0$, the integral includes the support of the Dirac delta, 
and we just get
the trivial identity that
$\Pi^{\rm gm}_{\eps=0} (\delta_0;\d;S)$ is equal to itself. However, 
if $\d-\delta_c> 0$, the
right-hand side vanishes and we get
$\Pi^{\rm gm}_{\eps=0} (\delta_0;\d;S)=0$. The same holds if $\d=\delta_c$. 
In this case only 
one half of the
support of $\Psi_{\eps}$ is inside the integration region,  so we get
$\Pi^{\rm gm}_{\eps=0} (\delta_0;\d;S)=
(1/2)\Pi^{\rm gm}_{\eps=0} (\delta_0;\d;S)$, 
which again 
implies $\Pi^{\rm gm}_{\eps=0} (\delta_0;\d;S)=0$. Therefore we
find that
\be\label{bc1}
\Pi^{\rm gm}_{\eps=0} (\delta_0;\d;S)=0
\hspace{10mm} {\rm if}\,\, \d\geq \delta_c\, .
\ee
This is not in contrast with the fact that 
$\Pi^{\rm gm}_{\eps}(\delta_0;\d;S)$ is the 
integral of the
positive definite quantity $W^{\rm gm}$. For finite $\eps$, 
$\Pi^{\rm gm} _{\eps}(\delta_0;\d;S)$ is indeed
strictly positive but, when $\d\geq \delta_c$, it vanishes  
in the limit $\eps\ra 0^+$.

Consider now \eq{CK3} when $\d<\delta_c$.  
In this case the zeroth-order term gives
a trivial identity. Pursuing the expansion to higher orders in
$\eps$
we have to take into account that
in $\Pi^{\rm gm}_{\eps} (\delta_0;\d;S+\eps)$ 
there is both an explicit dependence
on $\eps$ through the argument $S+\eps$, and a dependence implicit in the
subscript ${\eps}$. We begin by expanding the left-hand side as
\bees
\Pi^{\rm gm}_{\eps} (\delta_0;\d;S+\eps)&=&
\Pi^{\rm gm}_{\eps} (\delta_0;\d;S)
+\eps\frac{\pa\Pi^{\rm gm}_{\eps} (\delta_0;\d;S)}{\pa S}\nn\\
&&+\frac{\eps^2}{2}\frac{\pa^2\Pi^{\rm gm}_{\eps} (\delta_0;\d;S)}{\pa S^2}
+\ldots ,
\ees
without expanding for the moment the dependence on the index $\eps$.
On the right-hand side of \eq{CK3},  
we  expand $\Pi^{\rm gm}_{\eps} (\delta_0;\d-\D\d;S)$
in powers of $\D \d$,
\bees
&&\int_{\d-\delta_c}^{\infty} d(\D\d)\, \Psi_{\eps}(\D\d)
\Pi^{\rm gm}_{\eps} (\delta_0;\d-\D\d;t)\label{strictly0}\\
&=&\sum_{n=0}^{\infty}\frac{(-1)^n}{n!}
\frac{\pa^n\Pi^{\rm gm}_{\eps} (\delta_0;\d;S)}{\pa\d^n}\,
\int_{\d-\delta_c}^{\infty} d(\D\d)\, (\D\d)^n \Psi_{\eps}(\D\d)\, .\nn
\ees
Using \eq{Psi} we see that
\be\label{strictly}
\int_{\d-\delta_c}^{\infty} d(\D\d)\, (\D\d)^n \Psi_{\eps}(\D\d) \nn\\
=\frac{(2\eps)^{n/2}}{\sqrt{\pi}}\, 
\int_{-(\delta_c-\d)/\sqrt{2\eps}}^{\infty}dy\, y^n e^{-y^2}\, .
\ee
If $\d$ is {\em strictly} smaller 
than $\delta_c$ and $\delta_c-\d$ is finite (more precisely, if it does not scale with
$\sqrt{\eps}$) , the lower limit in the integration goes
to $-\infty$ as $\eps\ra 0^+$, and 
\bees
\int_{-(\delta_c-\d)/\sqrt{2\eps}}^{\infty}dy\, y^n e^{-y^2}&=&
\int_{-\infty}^{\infty}dy \, y^n e^{-y^2}
+{\cal O}\(e^{-(\delta_c-\d)^2/(2\eps)}\)\\
&&\hspace*{-8mm}=\frac{1+(-1)^n}{2}\, 
\frac{\sqrt{\pi}}{2^{n/2}}\, (n-1)!!+{\cal O}\(e^{-(\delta_c-\d)^2/(2\eps)}\)\, .\nn
\ees
The residue, being exponentially small in $\eps$, is beyond any order in the
expansion in powers of $\eps$, and we can neglect it, so
\be
\int_{\d-\delta_c}^{\infty} d(\D\d)\, (\D\d)^n \Psi_{\eps}(\D\d) \ra
\eps^{n/2}\,  (n-1)!!
\, ,
\ee
if $n$ even, and vanishes if $n$ is odd. 
Thus, \eq{CK3} gives 
\bees
&&\Pi^{\rm gm}_{\eps} (\delta_0;\d;S)
+\eps\frac{\pa\Pi^{\rm gm}_{\eps} (\delta_0;\d;S)}{\pa S}
+\frac{\eps^2}{2}\frac{\pa^2\Pi^{\rm gm}_{\eps} (\delta_0;\d;S)}{\pa S^2}
+\ldots \nn\\
&=&\Pi^{\rm gm}_{\eps} (\delta_0;\d;S)
+\frac{\eps}{2}\frac{\pa^2\Pi^{\rm gm}_{\eps} (\delta_0;\d;S)}{\pa\d^2}
+\frac{\eps^2}{8}\frac{\pa^4\Pi^{\rm gm}_{\eps} (\delta_0;\d;S)}{\pa\d^4}+\ldots
\nn\\\label{CK7}
\ees
From this structure it is clear that, when
$\delta_c-\d$ is finite, the dependence on the index $\eps$ in
$\Pi^{\rm gm}_{\eps}$ can be expanded in integer powers of $\eps$,
\be\label{CK8}
\Pi^{\rm gm}_{\eps} (\delta_0;\d;S) =\Pi^{\rm gm}_{\eps=0} (\delta_0;\d;S)
+\eps \Pi^{\rm gm}_{(1)} (\delta_0;\d;S) 
+\eps^2 \Pi^{\rm gm}_{(2)} (\delta_0;\d;S)
\ldots\, ,
\ee
where $\Pi^{\rm gm}_{(1)}$, $\Pi^{\rm gm}_{(2)}$, etc. are 
functions independent of $\eps$. 
We can now
collect the terms with the
same power of $\eps$ in the expansion of \eq{CK7}.  To order $\eps$ we find 
\be\label{FPxc0}
\frac{\pa\Pi^{\rm gm}_{\eps=0}(\delta_0;\d;S)}{\pa S}-
\frac{1}{2}\frac{\pa^2\Pi^{\rm gm}_{\eps=0}(\delta_0;\d;S)}{\pa\d^2}
= 0\, .
\ee
Putting together this result with \eq{bc1},
 we therefore end up with a FP equation with
the boundary condition 
$\Pi^{\rm gm}_{\eps=0}(\delta_0;\d=\delta_c;S)=0$, and therefore
we recover \eq{Pix0}. We have therefore succeeded in deriving the excursion
set formalism from our microscopic approach. Observe  that the
boundary condition $\Pi^{\rm gm}_{\eps=0}(\delta_0;\d=\delta_c;S)=0$ emerges
only when we take the continuum limit, and does not hold for finite $\eps$.

\subsection{Finite-$\eps$ corrections}\label{sect:finite_eps}

In  Section~\ref{sect:nonmark} we will find that the halo mass
function gets contributions, that we will call ``non-markovian'', that
depend on how $\Pi^{\rm gm}_{\eps}(\delta_0;\delta_c;S)$ 
approaches zero when $\eps\ra 0$.
It is therefore of great
importance for us to understand the finite-$\eps$ corrections to the
result obtained in the continuum limit. The issue is quite technical
and we summarize here the main results. Details are given
in appendix~\ref{app:A}.

As long as $\delta_c-\d$ is finite and strictly positive, we have seen that
the expansion (\ref{CK8}) applies, so the first correction to the
continuum result is ${\cal O}(\eps)$ and is given by $\eps\Pi^{\rm gm}_{(1)}$. 
Collecting  the next-to-leading terms in \eq{CK7}, 
we find that 
$\Pi^{\rm gm}_{(1)}$ satisfies a FP equation with the second time 
derivative of  $\Pi^{\rm gm}_{\eps=0}$  
as a source term,
\be\label{FPxc0eps2}
\frac{\pa\Pi^{\rm gm}_{(1)}(\delta_0;\d;S)}{\pa S}-
\frac{1}{2}\frac{\pa^2\Pi^{\rm gm}_{(1)}(\delta_0;\d;S)}{\pa\d^2}
= \frac{1}{4}\frac{\pa^2\Pi^{\rm gm}_{\eps =0}(\delta_0;\d;S)}{\pa S^2}
\, .
\ee
In the above derivation, a crucial point was that we  could extend
to $-\infty$ the lower integration limit in \eq{strictly}. This is
correct if we take the limit $\eps\ra 0^+$ with $\delta_c-\d$ fixed and
positive. The situation changes  at $\d=\delta_c$, since in this
case the lower limit of the integral is zero, rather than $-\infty$. 
In this case
\be
\int_0^{\infty}d(\D\d)\, (\D\d) \Psi_{\eps}(\D\d)=
\(\frac{\eps}{2\pi}\)^{1/2}\, ,
\ee
(while the same integral computed from $-\infty$ to $+\infty$
obviously vanished), so we now  have a term ${\cal O}\(\sqrt{\eps}\)$ on the
right-hand side of \eq{strictly0}. Furthermore,
\be
\int_0^{\infty}d(\D\d)\, \Psi_{\eps}(\D\d)=\frac{1}{2}\, ,
\ee
so the expansion of \eq{CK3} now gives
\bees
 \Pi^{\rm gm}_{\eps} (\delta_0;\delta_c;S)&=&
\frac{1}{2} \Pi^{\rm gm}_{\eps}(\delta_0;\delta_c;S)\label{nogood}\\
&&-\(\frac{\eps}{2\pi}\)^{1/2}
\left.\frac{\pa\Pi^{\rm gm}_{\eps}(\delta_0;\d;S)}{\pa\d}\right|_{x=\delta_c}
+\ldots\, .\nn
\ees
This indicates that
$\Pi^{\rm gm}_{\eps} (\delta_0;\delta_c;S)$ is  ${\cal O}(\eps^{1/2})$, rather
than ${\cal O}(\eps)$. However, \eq{nogood}
is not a good starting point for a quantitative
evaluation of $\Pi^{\rm gm}_{\eps} (\delta_0;\delta_c;S)$ since, as
we  show in appendix~\ref{app:A}, 
the expansion in derivatives becomes singular in
$\d=\delta_c$, and all terms denoted by the dots
in \eq{nogood} finally  give contributions of the same order in $\eps$.
A better procedure is the following. First, observe that the
correction is determined by the lower limit of the integral,
$(\delta_c-\d)/\sqrt{2\eps}$. The transition from the behavior ${\cal O}(\eps)$
valid for  $\delta_c-\d$ fixed and
positive, to the behavior ${\cal O}(\eps^{1/2})$ valid at  
$\d=\delta_c$ takes place
in a ``boundary layer'', consisting of the region where
$\delta_c-\d$ is positive and  ${\cal O}(\eps^{1/2})$, and the lower limit of the
integral is ${\cal O}(1)$. This is a situation that
often appears in stochastic processes near a boundary, or in fluid
dynamics, and can be treated by a standard technique (see
e.g. \cite{knessl}, where a very similar situation is discussed in terms
of the means first passage time, rather than in terms of the
distribution function $\Pi^{\rm gm}_{\eps}$). Namely, we introduce a 
``stretched variable'' $\eta$ (not to be confused, of course,
with the noise $\eta(t)$ of \eq{Langevin1})
\be
\eta =\frac{\delta_c-\d}{\sqrt{2\eps}}\, ,
\ee
which even as $\eps\ra 0^+$ is at most of order one inside the boundary layer, and we
write $\Pi^{\rm gm}_{\eps} (\delta_0;\d;S)$ in the form
\be\label{PiCu}
\Pi^{\rm gm}_{\eps} (\delta_0;\d;S)= C_{\eps} (\delta_0;\d;S)\, u(\eta)\, ,
\ee
where $C_{\eps} (\delta_0;\d;S)$ is a smooth function, while  the fast
variation inside the boundary layer is contained in $u(\eta)$.
By definition, we choose $u(\eta)$ such that
$\lim_{\eta\ra\infty} u(\eta) =1$,
so $C_{\eps}$ is just
the solution for $\Pi^{\rm gm}_{\eps}$ valid when
$\delta_c-\d$ is finite and positive, i.e.
$C_{\eps}$ is given by \eq{CK8}. Writing 
$\d=\delta_c-\eta\sqrt{2\eps}$  (and setting for
notational simplicity $\delta_0=0$) we have
\bees
&&C_{\eps} (\delta_0=0;\d;S)=\frac{1}{\sqrt{2\pi S}}\,\label{Ceps} \\
&&\times\[ \exp\left\{ -\frac{1}{2S}\(\delta_c-\eta\sqrt{2\eps}\)^2\right\}
-\exp\left\{ -\frac{1}{2S}\(\delta_c+\eta\sqrt{2\eps}\)^2\right\}
\]\, ,\nn
\ees
plus corrections ${\cal O}(\eps)$. Since $C_{\eps}$ by definition is smooth
everywhere, we can use \eq{Ceps} also inside the
boundary layer. In this case $\eta$ is at most 
${\cal O}(1)$, and we can expand the exponentials in \eq{Ceps} in powers of
$\sqrt{\eps}$.
In the limit $\eps\ra 0$,
\be
C_{\eps} (\delta_0=0;\d;S)=\sqrt{\eps}\, \frac{2\,\eta}{\sqrt{\pi}}\, 
\frac{\delta_c}{S^{3/2}} e^{-\delta_c^2/(2S)}\,  +{\cal O}(\eps)\, .
\ee
Plugging this result in \eq{PiCu} and sending $\d\ra \delta_c^-$ we find
\be\label{Pigammaeps}
\Pi^{\rm gm}_{\eps} (\delta_0;\delta_c;S)=
\sqrt{\eps}\,\, \gamma
\frac{\delta_c}{S^{3/2}} e^{-\delta_c^2/(2S)} +{\cal O}(\eps)\, ,
\ee
where
\be\label{defgamma}
\gamma = \frac{2}{\sqrt{\pi}}\,\lim_{\eta\ra 0}\eta\, u(\eta)\, .
\ee
In appendix~\ref{app:A} we show that $\g=1/\sqrt{\pi}$, so
\be\label{Pigammafinal}
\Pi^{\rm gm}_{\eps} (\delta_0;\delta_c;S)=
\sqrt{\eps}\, \frac{1}{\sqrt{\pi}}\, 
\frac{\delta_c-\delta_0}{S^{3/2}} e^{-(\delta_c-\delta_0)^2/(2S)} +{\cal O}(\eps)\, .
\ee
Similarly, for $\delta_n<\delta_c$,
\be\label{Pigammafinalbis}
\Pi^{\rm gm}_{\eps} (\delta_c;\delta_n;S)=
\sqrt{\eps}\, \frac{1}{\sqrt{\pi}}\, 
\frac{\delta_c-\delta_n}{S^{3/2}} e^{-(\delta_c-\delta_n)^2/(2S)} +{\cal O}(\eps)\, .
\ee
Observe that at the numerator of \eqs{Pigammafinal}{Pigammafinalbis}
always enters the absolute value of the
difference of the first two arguments of $\Pi^{\rm gm}_{\eps}$, i.e. 
$\delta_c-\delta_0$ in \eq{Pigammafinal} and $\delta_c-\delta_n$
in \eq{Pigammafinalbis}, as it is also obvious from the fact that
$\Pi^{\rm gm}_{\eps}$ is definite positive.
\Eqs{Pigammafinal}{Pigammafinalbis}
will be important when we compute the non-markovian
corrections, in Section~\ref{sect:nonmark}.
To conclude this section, it is interesting to discuss the behavior of
$\Pi^{\rm gm}_{\eps} (\delta_0;\d;S)$ for $\d$ {\em larger} than
$\delta_c$, with $\d-\delta_c$ finite (and, as always in this
section, $\delta_0<\delta_c$). In this case the lower integration limit in
\eq{strictly} goes to $+\infty$ as $\eps\ra 0^+$ and 
\be
\Pi^{\rm gm}_{\eps} (\delta_0;\d;S)\sim
\frac{1}{\sqrt{2\pi\eps}}\, \exp\{-(\delta_c-\d)^2/(2\eps)\}\, .
\ee
This function is zero to all orders in a Taylor expansion around 
$\eps=0^+$.

\section{Extension of  excursion  set theory to
generic filter}\label{sect:extensiongau}

We next consider 
the computation of  the distribution function
$\Pi_{\eps} $, still restricting for the moment
to gaussian fluctuations, but using a
generic filter function. In this case 
the natural time variable is the  variance $S$ 
computed with the chosen filter function, so  in the following $S$ 
denotes the variance computed with the filter function that one
is considering. Again
we discretize it in equally spaced steps,
$S_k=k\eps$,
with $S_n=n\eps \equiv S$, and a trajectory is  defined by 
the collection of values $\{\delta_1,\ldots ,\delta_n\}$, 
such that $\delta(S_k)=\delta_k$.

The distribution function for gaussian fluctuations and arbitrary
filter function is given by \eq{3defPi0xiixij}.
As we saw in the previous section, in the markovian case
$\Pi_{\eps}$ satisfies a local differential equation,
namely the Fokker-Planck equation.
It is instructive to understand
that, for a generic filter, it is no longer possible to 
write a local
diffusion equation for $\Pi_{\eps}(\delta_0;\delta_n;S_n)$. This will
immediately make it clear that the problem is now significantly more complex. 
Indeed, by taking the derivative with respect to 
$S_n$ of both sides of eq. (\ref{3defPi0xiixij}), we get
\bees
&&\hspace*{-5mm}\frac{\pa}{\pa S_n}\Pi_{\eps}(\delta_0;\delta_n;S_n) = 
\frac{1}{2}\sum_{k,l=1}^n \frac{\pa\langle\delta_k\delta_l\rangle_c}{\pa S_n}
\label{noFP}\\
&&\times\int_{-\infty}^{\delta_c} 
d\delta_1\ldots d\delta_{n-1}\,\pa_k\pa_l 
W(\delta_0;\delta_1,\ldots ,\delta_n;S_n)\, ,\nn
\ees
where $\pa_k\equiv\pa/\pa\d_k$, and we used the fact that, acting on $\exp\{i\sum_{i=1}^n\lambda_i \delta_i\}$, 
$\pa_k$ gives $i\lambda_k$.
Therefore,
separating the term with $k=l=n$ from the rest, and observing that
$\langle\delta(S_k)\delta(S_l)\rangle_c$ depends on $S_n$ only if at least
one of the two indices $k$ or $l$ is equal to $n$, we get
\bees
&&\frac{\pa}{\pa S_n}\Pi_{\eps}(\delta_0;\delta_n;S_n) = 
\frac{1}{2}\frac{\pa^2}{\pa \delta_n^2}\Pi_{\eps}(\delta_0;\delta_n;S_n)\label{eqnonloc}\\
&&+\sum_{k=1}^{n-1} \frac{\pa\langle\delta_k\delta_n\rangle_c}{\pa S_n}
\pa_n\int_{-\infty}^{\delta_c} 
d\delta_1\ldots d\delta_{n-1}\,\pa_k
W(\delta_0;\delta_1,\ldots ,\delta_n;S_n).\nn
\ees
If the upper limit of the integrals were $+\infty$, rather than $\delta_c$, 
the term proportional to $\pa_kW$ with $k<n$
would give zero, since it is a total derivative with respect to one
of the integration variables $d\delta_1, \ldots d\delta_{n-1}$, and $W$ vanishes
exponentially when any of its arguments $\delta_k$ goes to
$\pm\infty$. Thus, one would remain with a Fokker-Planck
equation. However, when the upper limit $\delta_c$ is finite,
the terms proportional to $\pa_kW$ with $k<n$
give in general non-vanishing boundary
term. Actually, for a sharp $k$-space filter, we found that
$\langle\delta_k\delta_n\rangle_c={\rm min}(S_k,S_n)=S_k$, which is
independent of $S_n$ for $k<n$. Therefore
$\pa\langle\delta_k\delta_n\rangle_c/\pa S_n=0$, 
and the term in the second line
of \eq{eqnonloc} vanishes. This is another way of showing that,
in the continuum limit, for sharp $k$-space filter the probability
distribution satisfies a FP equation, as we
already found in Section~\ref{subsect:cont}.\footnote{Note however that
this only holds  in the continuum limit, as is implicit in the fact that we
are taking the derivative with respect to $S_n$, which means that we
are considering $S_n$ has a continuous variable.} 

For a generic form of the two-point correlator,
the term  in the second line of \eq{eqnonloc}
is non-vanishing, and in general it is very complicated. Furthermore,
in the continuum limit the sum over $k$ in \eq{eqnonloc} becomes an
integral over an intermediate time variable $S_k$, so this term is
non-local with respect to "time" $S$.
Thus, we can no longer 
determine
$\Pi_{\eps}(\delta_0;\delta_n;S_n)$ by solving a local
differential equation, as we did in the markovian case. 
Once again, this shows that the common procedure of using the
distribution function computed with the $k$-space filter, and
substituting in it the relation between
mass and smoothing radius  of the tophat
filter in coordinate space, is not justified. What we need is to formulate the problem in such a way that
it becomes possible to treat the non-markovian terms as perturbations,
which is not at all evident from \eq{eqnonloc}.

In this section we  develop such a perturbative scheme.
We illustrate the computation of $\Pi_{\eps}(\delta_0;\delta_n;S_n)$
using a tophat filter in coordinate space,
which is finally the most interesting case since we can associate to it 
a well defined
mass, but the technique that we  develop can be used more
generally.

In  Section~\ref{sect:twopointcorr}
we study  the two-point correlator 
with tophat filter in coordinate space and we show
that it can be split into two parts, which we  call
markovian and non-markovian, respectively. 
In Section~\ref{subsect:marterm}  we  compute  the contribution
of the markovian term  to the halo mass function, while 
in Section~\ref{sect:nonmark}
we  develop the formalism for computing perturbatively the contribution 
of the non-markovian term.

\subsection{The two-point correlator 
with tophat filter in coordinate space}\label{sect:twopointcorr}

We first  study  the
correlator 
$\langle\d(R_1)\d(R_2)\rangle$ with a tophat filter in coordinate space. 
We use \eq{dxdkWk} with ${\bf x}=0$. 
The two-point correlator of the  non-smoothed
density contrast is given in \eq{defP(k)}. We write the power spectrum
after recombination 
as $P(k)T^2(k)$, where $P(k)$ is the primordial power spectrum
and $T(k)$ is the transfer function, 
so for the smoothed density contrast we get
\be\label{2pointWW}
\langle\d(R_1)\d(R_2)\rangle=
\frac{1}{2\pi^2}\int_0^{\infty} dk\, k^2P(k)T^2(k)
\tilde{W}(k,R_1)\tilde{W}^*(k,R_2)\, .
\ee

\begin{figure}
\includegraphics[width=0.45\textwidth]{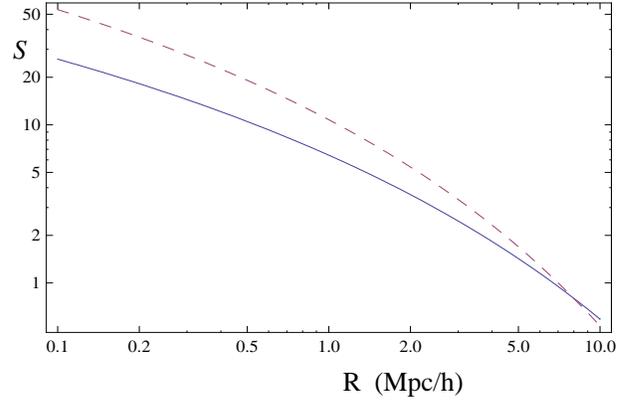}
\caption{\label{fig:mu2_and_S_vs_R}
The functions $S(R)$ computed for a tophat filter in coordinate
space
(blue solid curve) and $S(R)$ computed for sharp $k$-space filter
(violet, dashed), against $R$, on a log-log scale.
}
\end{figure}

\noindent
When $R_1=R_2=R$, this  reduces to  $S(R)$.
We consider a   primordial spectrum $P(k)=A k^{n_s}$,
processed into the post-recombination spectrum by the transfer 
function $T(k)$ as in \cite{Sugiyama}, in a concordance $\Lambda$CDM model
with a power spectrum normalization $\s_8=0.8$
and 
$h=0.7$, $\Omega_M=1-\Omega_{\Lambda}=0.28$, $\Omega_B=0.046$
and $n_s=0.96$,
consistent with the WMAP 5-years data release. 

We first study  $S(R)$.
We compute the integral in \eq{mu2RW2}
numerically, for different values of $R$,
both with the sharp $k$-space filter (\ref{Wsharpk}) with
$k_f=1/R$, and with the tophat filter in coordinate $x$-space, whose Fourier
transform is
\be\label{Wtildesharpx}
\tilde{W}_{{\rm sharp}-x}(k,R)=3\,
\frac{\sin(kR)-kR\cos(kR)}{(kR)^3}\, .
\ee
For both filters, the constant $A$ in $P(k)$ is fixed so that
$S=\s_8$ when $R=(8/h)$~Mpc. The result is shown
in Fig.~\ref{fig:mu2_and_S_vs_R}.

\begin{figure}
\includegraphics[width=0.45\textwidth]{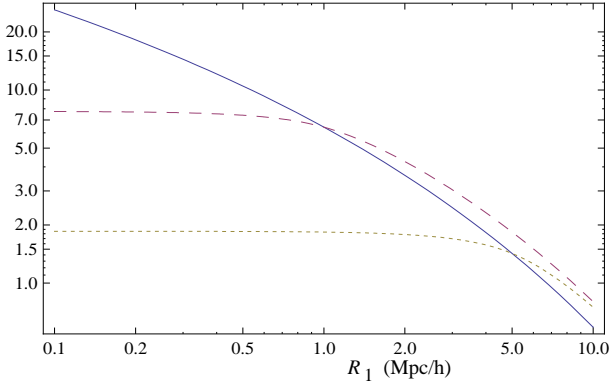}
\caption{\label{fig:mu2R1R2}
The quantity $S(R_1)$ for a tophat filter in coordinate space
(blue solid curve),  the correlator $\langle\d(R_1)\d(R_2)\rangle$ with
$R_2=1 {\rm Mpc/h}$ (violet dashed line) and
$\langle\d(R_1)\d(R_2)\rangle$ with
$R_2=5 {\rm Mpc/h}$ (brown dotted line), as functions of $R_1$. 
}
\end{figure}

We consider next the correlator (\ref{2pointWW}) with
the tophat filter in coordinate space. We compute the integral
in \eq{2pointWW}
numerically, holding $R_2$ fixed and varying $R_1$. The result is
shown in Fig.~\ref{fig:mu2R1R2}. The solid line is the function 
$S(R_1)$,
already
shown
in Fig.~\ref{fig:mu2_and_S_vs_R}.
The dashed line is $\langle\d(R_1)\d(R_2)\rangle$ with
$R_2=1\, {\rm Mpc}/h$, as a function
of $R_1$, while the dotted line is $\langle\d(R_1)\d(R_2)\rangle$ with
$R_2=5\, {\rm Mpc}/h$, again as a function
of $R_1$. We see that, as long as $R_1<R_2$, the two-point correlator is 
approximately constant and equal to $S(R_2)$, while for
$R_1>R_2$ the correlator is approximately equal to  $S(R_1)$.
In other words,
\be\label{appro}
\langle\d(R_1)\d(R_2)\rangle\simeq 
{\rm min}(S(R_1),S(R_2))\, .
\ee

\begin{figure}
\includegraphics[width=0.45\textwidth]{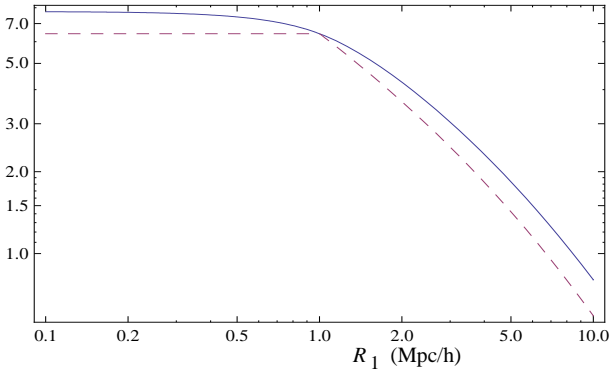}
\caption{\label{fig:minR1R2}
The  correlator $\langle\d(R_1)\d(R_2)\rangle$ (blue solid line), compared to
${\rm min}(S(R_1),S(R_2))$ (violet, dashed). In both cases 
$R_2=1 {\rm Mpc/h}$, and we plot the functions against $R_1$,
measured in ${\rm Mpc}/h$.
}
\end{figure}

In  Fig.~\ref{fig:minR1R2} we compare
$\langle\d(R_1)\d(R_2)\rangle$ (blue solid line) and
${\rm min}(S(R_1),S(R_2))$ (violet dashed line).
This result suggests to define a function $C(R_1,R_2)$ from
\be
\langle\d(R_1)\d(R_2)\rangle = 
{\rm min}(S(R_1),S(R_2)) +C(R_1,R_2)\, .
\ee
As we
see from Fig.~\ref{fig:mu2_and_S_vs_R}, the function $S(R)$ can be inverted
to give $R=R(S)$, so $R_1=R(S_1)$ and $R_2=R(S_2)$. We define
\be
\D(S_1,S_2)=C(R(S_1),R(S_2))\, ,
\ee
and we can write
\be\label{rever}
\langle\delta_i\delta_j\rangle = 
{\rm min}(S_i,S_j) + \D(S_i,S_j)\, .
\ee
If one simply
neglects $\D(S_i,S_j)$, i.e. one makes the approximation (\ref{appro}),
the problems become formally identical to the one that we have solved
in Section~\ref{sect:Bond}. Therefore,
we end up with the standard excursion set theory result given
in \eq{PiChandra}, and therefore with the PS mass function,
in which $S$ is simply the  variance computed with
the filter of our choice, in this case the tophat filter in coordinate
space. The corrections to this result are due
to $\D(S_i,S_j)$, so it is useful first of all to
understand better the form of this function.

\begin{figure}
\includegraphics[width=0.45\textwidth]{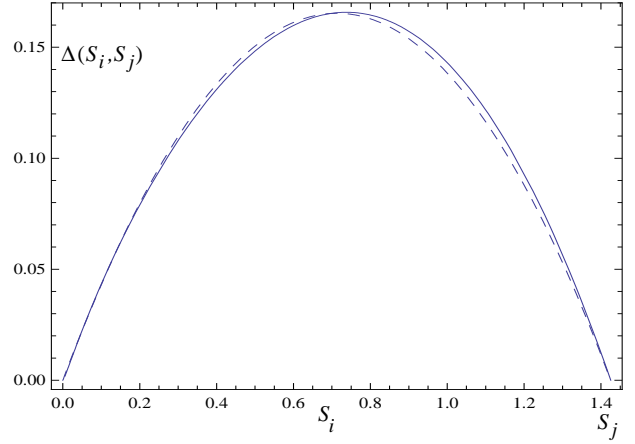}
\caption{\label{fig:C2TiTn}
The function  $\D(S_i,S_j) $ for tophat
filter in coordinate space (solid line),
with $S_j\simeq 1.42$ (corresponding to $R(S_j)=
5~{\rm Mpc}/h$),  plotted against $S_i$, in the range $0\leq S_i\leq S_j$,
and the function $\kappa S_i(S_j-S_i)/S_j$ with
$\kappa\simeq 0.45$ (dashed line). 
}
\end{figure}

By definition $\D(S_i,S_j)$ is symmetric, $\D(S_i,S_j)=\D(S_j,S_i)$, so
it is sufficient to study it for $S_i\leq S_j$. 
We also use the notation $\D_{ij}=\D(S_i,S_j)$. 
Since, by definition, $\langle\delta_i^2\rangle =S_i$, we see from
\eq{rever} that $\D(S_i,S_j)$ vanishes when $S_i=S_j$.
Furthermore, at $S_i=0$, $\delta_i=\delta_0$ is the same constant  for all
trajectories,  so $\langle\delta_i\delta_j\rangle_c
=\delta_0\langle\delta_j\rangle_c=0$, and therefore $\D(S_i,S_j)$ vanishes
when $S_i=0$.

In Fig.~\ref{fig:C2TiTn} we plot $\D(S_i,S_j)$ for $S_j$ fixed, as a
function of $S_i$, with $0\leq S_i\leq S_j$, for our reference 
$\Lambda$CDM model
(solid line). The dashed line  in
Fig.~\ref{fig:C2TiTn} is the approximation
\be\label{CTTaT}\label{approxDelta}
\D(S_i,S_j)\simeq \kappa\, \frac{S_i(S_j-S_i)}{S_j}\, ,
\ee
with $\kappa\simeq 0.45$ (a more accurate value will be given below). 
We see that
\eq{CTTaT} provides an 
excellent analytical approximation 
to $\D(S_i,S_j)$.\footnote{Varying $S_j$ we find that \eq{CTTaT} 
becomes exact
  (within our numerical accuracy) for small $S_j$, while for large
  $S_j$ the function $S_i (S_j-S_i)/S_j$ must be replaced by a a less
  symmetric function, whose maximum is at a value of $S_i$
slightly larger than
  $S_j/2$. The qualitative shape of the function remains however the
same.
For completeness, we have also considered  a gaussian
filter. In this case we find that, in a first approximation, the
function $\D_{ij}$ is still given by \eq{CTTaT} (although the
actual form of $\D_{ij}$ is slightly more skewed compared to an inverse
parabola),  with a value
of $\kappa\simeq 0.35$. 
}

For $S_j$ fixed and $S_i\ra 0$, 
the correction $\D(S_i,S_j)$ is linear
in $S_i$,  so more generally  we can define $\kappa(S_j)$ from 
$\kappa(S_j)=\lim_{S_i\ra 0} \D(S_i,S_j)/S_i$, or equivalently,
\be\label{adiRlim}
\kappa (R) = 
\lim_{R'\ra \infty}
\frac{\langle\d(R')\d(R)\rangle}{\langle\d^2(R')\rangle} -1\, .
\ee
In the $\Lambda$CDM model that we are using, 
our numerical results  display a very weak linear dependence of $\kappa$
on $R$. Taking for instance the data in the range
$R\in [1,60] \, {\rm Mpc}/h$, the result of the numerical evaluation
of \eq{adiRlim}
is very well
fitted by 
\be\label{Ar}
\kappa (R) \simeq 0.4592-0.0031\, R\, ,
\ee
where $R$ is measured in ${\rm Mpc}/h$.\footnote{The value of $\kappa$
  depends in principle on the cosmological model used, but this dependence
  is quite weak. For comparison,
  using a $\Lambda$CDM cosmological model with $h=0.7$, 
$\Omega_M=1-\Omega_{\Lambda}=0.3$, $\s_8=0.93$, $\Omega_Bh^2=0.022$
  and $n_s=1$, consistent with the WMAP 1st year data release, gives
$\kappa (R) \simeq 0.4562-0.0040\, R$.}

\subsection{Markovian term}\label{subsect:marterm}

Inserting \eq{rever} into \eq{3defPi0xiixij} we get
\bees
&&\Pi_{\eps}(\delta_0;\delta_n;S_n) =
\int_{-\infty}^{\delta_c} 
d\delta_1\ldots d\delta_{n-1}\,\Dl\label{Delta1}\\
&&\times\exp\left\{
i\sum_{i=1}^n\lambda_i\delta_i -\frac{1}{2}\sum_{i,j=1}^n
[{\rm min}(S_i,S_j) + \D(S_i,S_j)]
\lambda_i\lambda_j\right\}\nn\, .
\ees
As we see from Fig.~\ref{fig:minR1R2},
\eq{appro} gives a reasonable approximation to the exact correlator.
This suggests to treat $\D_{ij}$ as a perturbation, so
we now expand  in $\D_{ij}$.
The zeroth-order term is simply
$\Pi^{\rm gm}_{\eps}(\delta_0;\delta_n;S_n)$, whose continuum limit is given in 
\eq{Pix0}. The corresponding first-crossing rate is
\bees
{\cal F}^{\rm gm}&=&-\int_{-\infty}^{\delta_c}d\delta\,
\frac{\pa\Pi^{\rm gm}_{\eps=0}}{\pa S}\nn\\
&=&\frac{1}{\sqrt{2\pi}}\, \frac{\delta_c}{S^{3/2}}\, e^{-\delta_c^2/(2S)}
\label{Fmarkov}\, .
\ees
so
the markovian term can be obtained by taking the excursion set result
(\ref{Pix0}), which was computed with the sharp $k$-space filter, and
replacing the variance computed with 
the sharp $k$-space filter with the variance computed
with the filter of interest. This is the procedure that is normally
used in the literature. From our vantage point, we now see that the
corrections to this procedure are given by the non-markovian
contributions, to which we now turn.

\subsection{Non-markovian corrections}\label{sect:nonmark}

We now discuss
the non-markovian corrections, to first order,
using the analytical approximation (\ref{approxDelta}) for $\D_{ij}$.
From \eq{Delta1}, expanding to first order in $\D_{ij}$
and using
$\lambda_ie^{i\sum_k\lambda_k \d_k}=-i\pa_ie^{i\sum_k\lambda_k\d_k}$, 
where $\pa_i=\pa/\pa\d_i$, the first-order correction to
$\Pi_{\eps}$ is
\bees
&&\Pi^{{\D}1}_{\eps}(\delta_0;\delta_n;S_n) \equiv
\int_{-\infty}^{\delta_c} 
d\delta_1\ldots d\delta_{n-1}\,\frac{1}{2}\sum_{i,j=1}^n \D_{ij}\pa_i\pa_j\nn\\
&&\times
\int{\cal D}\lambda\, 
\exp\left\{
i\sum_{i=1}^n\lambda_i\delta_i -\frac{1}{2}\,\sum_{i,j=1}^n {\rm min}(S_i,S_j)
\lambda_i\lambda_j\right\}\label{Delta3}\\
&&=\frac{1}{2}\sum_{i,j=1}^n \D_{ij}\int_{-\infty}^{\delta_c} 
d\delta_1\ldots d\delta_{n-1}\,\pa_i\pa_j
W^{\rm gm}(\delta_0;\delta_1,\ldots ,\delta_n;S_n)
\, .\nn
\ees
We rewrite the term $ \D_{ij}\pa_i\pa_j$ separating
explicitly the derivative $\pa_n\equiv \pa/\pa \delta_n$
from the derivatives $\pa_i$ with
$i<n$, so (using  $\D_{ij}=\D_{ji}$)
\be\label{Dsum}
\frac{1}{2} 
\sum_{i,j=1}^n \D_{ij}\pa_i\pa_j
= \frac{1}{2} \D_{nn}\pa_n^2 +\sum_{i=1}^{n-1}\D_{in}\pa_i\pa_n+
\frac{1}{2}\sum_{i,j=1}^{n-1}\D_{ij}\pa_i\pa_j\, .
\ee
Since $\D_{ij}=0$ when $i=j$, 
the above equation simplifies to
\be
\frac{1}{2} 
\sum_{i,j=1}^n \D_{ij}\pa_i\pa_j
=\sum_{i=1}^{n-1}\D_{in}\pa_i\pa_n+
\sum_{i<j}\D_{ij}\pa_i\pa_j\, ,\nn
\ee
where 
\be
\sum_{i<j}\equiv \sum_{i=1}^{n-2}\sum_{j=i+1}^{n-1}\, .
\ee
When inserted into \eq{Delta3}
the term
$\sum_{i=1}^{n-1}\D_{in}\pa_i\pa_n$ brings a factor $\sum_i$ that, in
the continuum limit, produces an integral over an intermediate
time $S_i$. Because of this dependence on the past history, we call
this the ``memory term''. Similarly, the term
$\sum_{i<j}\D_{ij}\pa_i\pa_j$ gives, in the
continuum limit, a double integral over intermediate times $S_i$ and $S_j$,
and we call it the ``memory-of-memory'' term.  Thus,
\be\label{PiD1}
\Pi^{{\D}1}_{\eps}=
\Pi_{\eps}^{\rm mem}+\Pi_{\eps}^{\rm mem-mem}\, ,
\ee
where
\bees
&&\Pi_{\eps}^{\rm mem}(\delta_0;\delta_n;S_n)\label{defPimemory}\\
&&=\sum_{i=1}^{n-1}\D_{in}\pa_n
\int_{-\infty}^{\delta_c} 
d\delta_1\ldots d\delta_{n-1}\,\pa_iW^{\rm gm}(\delta_0;\delta_1,\ldots ,\delta_n;S_n)\, ,\nn
\ees
and
\bees
&&\Pi_{\eps}^{\rm mem-mem}(\delta_0;\delta_n;S_n)\label{defPimem-memory}\\
&&=\sum_{i<j}\D_{ij}
\int_{-\infty}^{\delta_c} 
d\delta_1\ldots d\delta_{n-1} \pa_i\pa_j\,W^{\rm gm}(\delta_0;\delta_1,\ldots ,\delta_n;S_n)
\, .\nn
\ees
If we expand to quadratic and higher orders in $\D_{ij}$, we get terms with a
higher and higher number of summations
(or, in the continuum limit, of integrations) 
over intermediate time variables.

To compute the memory term we  integrate $\pa_i$ by parts,
\bees
&&\int_{-\infty}^{\delta_c} 
d\delta_1\ldots d\delta_{n-1}\,\pa_iW^{\rm gm}(\delta_0;\delta_1,\ldots ,\delta_n;S_n)\\
&&=\int_{-\infty}^{\delta_c} d\delta_1\ldots \widehat{d\d}_i\ldots 
d\delta_{n-1}
W(\delta_0;\delta_1,\ldots, \delta_i=\delta_c,\ldots ,\delta_{n-1},\delta_n;S_n)\, ,\nn
\ees
where
the notation $ \widehat{d\d}_i$ means that we must omit $d\delta_i$ from
the list of integration variables. 
We next observe that, because of
the property (\ref{markov}), $W^{\rm gm}$ satisfies
\bees\label{facto}
&&W^{\rm gm}(\delta_0;\delta_1,\ldots, \delta_{i-1}, \delta_c, \delta_{i+1}, \ldots ,\delta_n; S_n)\\
&&= 
W^{\rm gm}(\delta_0;\delta_1,\ldots, \delta_{i-1}, \delta_c;S_i)
W^{\rm gm}(\delta_c; \delta_{i+1}, \ldots ,\delta_n; S_n-S_i)\, ,\nn
\ees
so
\bees\label{WPiPi}
&&\int_{-\infty}^{\delta_c}d\delta_1\ldots d\delta_{i-1}
\int_{-\infty}^{\delta_c}  d\delta_{i+1}\ldots  d\delta_{n-1}\nn\\
&&\times W^{\rm gm}(\delta_0;\delta_1,\ldots, \delta_{i-1}, \delta_c;S_i)
W^{\rm gm}(\delta_c; \delta_{i+1}, \ldots ,\delta_n; S_n-S_i)\nn\\
&&= \Pi^{\rm gm}_{\eps}(\delta_0;\delta_c;S_i)
\Pi^{\rm gm}_{\eps}(\delta_c;\delta_n;S_n-S_i)\, ,
\ees
and we get
\be\label{Pimem1}
\Pi_{\eps}^{\rm mem}(\delta_0;\delta_n;S_n)=\sum_{i=1}^{n-1}\D_{in}\pa_n
\[ \Pi^{\rm gm}_{\eps}(\delta_0;\delta_c;S_i)
\Pi^{\rm gm}_{\eps}(\delta_c;\delta_n;S_n-S_i)\] .
\ee
In the continuum limit we  write
\be\label{sumint}
\sum_{i=1}^{n-1}\ra \frac{1}{\eps}\int_0^{S_n} dS_i\, ,
\ee
and, using \eqs{Pigammafinal}{Pigammafinalbis}, we find
\bees\label{Pimem2}
\Pi_{\eps=0}^{\rm mem}(\delta_0=0;\delta_n;S_n)&=&\frac{1}{\pi}
\pa_n\int_0^{S_n}dS_i\,
\D(S_i,S_n)\frac{\delta_c(\delta_c-\delta_n)}{S_i^{3/2}(S_n-S_i)^{3/2}}
\nn\\
&&\times
\exp\left\{-\frac{\delta_c^2}{2S_i}-\frac{(\delta_c-\delta_n)^2}{2(S_n-S_i)}
\right\}\, .
\ees
We now insert the form (\ref{approxDelta}) for $\D_{ij}$. The integral
can be computed exactly using the identities
\bees
&&\int_0^{S_n}dS_i\, \frac{S_i}{S_i^{3/2} (S_n-S_i)^{3/2}}\,
\exp\left\{-\frac{a^2}{2S_i}-\frac{b^2}{2(S_n-S_i)}\right\}\nn\\
&=& \sqrt{2\pi}\,\, \frac{1}{b}\, \frac{1}{S_n^{1/2}}
\exp\left\{-\frac{(a+b)^2}{2S_n}\right\}\, ,
\label{magic1}
\ees
and
\bees
&&\int_0^{S_n}dS_i\, \frac{S_i^2}{S_i^{3/2} (S_n-S_i)^{3/2}}\,
\exp\left\{-\frac{a^2}{2S_i}-\frac{b^2}{2(S_n-S_i)}\right\}\nn\\
&=& \sqrt{2\pi}\,\, \frac{S_n^{1/2}}{b}\, 
\exp\left\{-\frac{(a+b)^2}{2S_n}\right\}
-\pi\, {\rm Erfc}\(\frac{a+b}{\sqrt{2S_n}}\)
\, ,
\label{magic2}
\ees
where ${\rm Erfc}$ is the complementary error 
function.\footnote{To derive these results we  take one derivative
of the left-hand side of \eq{magic1}  with respect to $a^2$. The
resulting integral can be performed using \eq{magic}, and we then
integrate back with respect to $a^2$. Similarly, \eq{magic2} is
obtained taking twice the derivative with respect to $a^2$.}
This gives
\be\label{Pimempan}
\Pi_{\eps=0}^{\rm mem}(\delta_0=0;\delta_n;S_n)
=\kappa \pa_n \[ \frac{\delta_c(\delta_c-\delta_n)}{S_n}\, 
{\rm Erfc}\(\frac{2\delta_c-\delta_n}{\sqrt{2S_n}}\)\]\, .
\ee
For the memory-of-memory term, proceeding as for the memory term, we get

\begin{figure}
\includegraphics[width=0.45\textwidth]{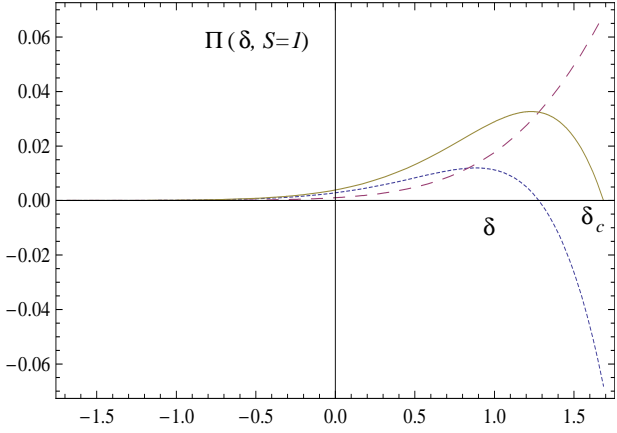}
\caption{\label{fig:Pimemmem1}
The functions  $\Pi_{\eps=0}^{\rm mem}(\delta_0=0;\d;S=1)$ (blue dotted line)
$\Pi_{\eps=0}^{\rm mem-mem}(\delta_0=0;\d;S=1)$ (violet, dashed) and their sum
(brown solid line), as functions of $\d$.
}
\end{figure}

\begin{figure}
\includegraphics[width=0.45\textwidth]{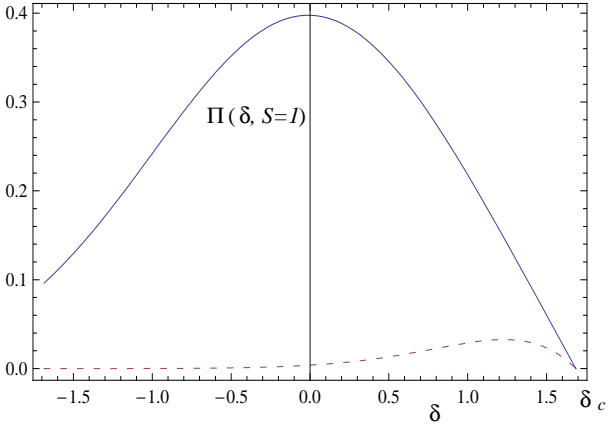}
\caption{\label{fig:Pimemmem2}
The functions  $\Pi_{\eps=0}^{\rm mem}(\delta_0=0;\d;S=1)+
\Pi_{\eps=0}^{\rm mem-mem}(\delta_0=0;\d;S=1)$ (dashed),
compared to $\Pi_{\eps=0}^{\rm gm}(\delta_0=0;\d;S=1)$
(solid line), as functions of $\d$.
}
\end{figure}

\bees
&&\Pi_{\eps}^{\rm mem-mem}(\delta_0;\delta_n;S_n)\label{Pimemmem1}\\
&=&\sum_{i<j}\D_{ij}
\Pi^{\rm gm}_{\eps}(\delta_0;\delta_c;S_i)
\Pi^{\rm gm}_{\eps}(\delta_c;\delta_c;S_j-S_i)
\Pi^{\rm gm}_{\eps}(\delta_c;\delta_n;S_n-S_j)\, .\nn
\ees
To compute this quantity we also  need
$\Pi^{\rm gm}_{\eps}(\delta_c;\delta_c;S)$, with both the first  and the
second arguments equal to $\delta_c$. As we discuss in
appendix~\ref{app:A}, the 
result is
\be\label{Pixcxc}
\Pi^{\rm gm}_{\eps}(\delta_c;\delta_c;S)=
\frac{\eps}{\sqrt{2\pi}\,\, S^{3/2}}\, .
\ee
Actually, \eq{Pixcxc} is exact, and not just valid to
${\cal O}(\eps)$.
Using \eqs{Pigammafinal}{Pixcxc} we get
\bees
&&\Pi_{\eps=0}^{\rm mem-mem}(\delta_0=0;\delta_n;S_n)=
\frac{\kappa}{\pi\sqrt{2\pi}}\, \delta_c(\delta_c-\delta_n)\nn\\
&&\times
\int_0^{S_n}dS_i\, \frac{1}{S_i^{1/2}}\, e^{-\delta_c^2/(2S_i)}
\\
&&\times\int_{S_i}^{S_n}dS_j
\frac{1}{S_j(S_j-S_i)^{1/2}(S_n-S_j)^{3/2} }
\exp\left\{-\frac{(\delta_c-\delta_n)^2}{2(S_n-S_j)}\right\}\, .\nn
\ees
It is convenient to
use the identity
\be\label{ratherthan}
\frac{(\delta_c-\delta_n)}{(S_n-S_j)}
\exp\left\{-\frac{(\delta_c-\delta_n)^2}{2(S_n-S_j)}\right\}
=\pa_n \exp\left\{-\frac{(\delta_c-\delta_n)^2}{2(S_n-S_j)}\right\}\, ,
\ee
to write $\Pi_{\eps=0}^{\rm mem-mem}$ as a total derivative with respect
to $\delta_n$.
The inner integral can now be computed rewriting it in terms of the variable
$z=(\delta_c-\delta_n)^2/[2(S_n-S_j)]$, and gives
\bees
&&\hspace*{-5mm}\Pi_{\eps=0}^{\rm mem-mem}(\delta_0=0;\delta_n;S_n)=
\frac{\kappa \delta_c}{\sqrt{2\pi S_n}}\,
\pa_n\[ e^{-(\delta_c-\delta_n)^2/(2S_n)}\right.\nn\\
&&\hspace*{-5mm}\times \left.\int_0^{S_n} \frac{dS_i}{S_i}\, e^{-\delta_c^2/(2S_i)}
\,{\rm Erfc}\((\delta_c-\delta_n)\sqrt{\frac{S_i}{2(S_n-S_i) S_n}}\,\)\]
.\label{Pimemmempan}
\ees
We have not been able to compute analytically this last integral, but
the fact that $\Pi_{\eps=0}^{\rm mem-mem}$ is a total derivative with
respect to $\delta_n$ will allow us to compute analytically the
first-crossing rate, see below.
First, it is interesting to plot the functions 
$\Pi_{\eps=0}^{\rm mem}$
and $\Pi_{\eps=0}^{\rm mem-mem}$. We show them in
Fig.~\ref{fig:Pimemmem1}, setting for definiteness $S_n=1$.
Observe that these two functions are separately
non-zero in $\d=\delta_c$. However, 
\be
\Pi_{\eps=0}^{\rm mem}(\delta_0=0;\delta_c,S_n)=-\frac{\kappa \delta_c}{S_n}
{\rm Erfc}\(\frac{\delta_c}{\sqrt{2 S_n}}\)\, ,
\ee
and
$\Pi_{\eps=0}^{\rm mem-mem}(\delta_0=0;\delta_c,S_n)=
-\Pi_{\eps=0}^{\rm mem}(\delta_0=0;\delta_c,S_n)$,
so we find that
the total distribution function $\Pi_{\eps=0}(\delta_0;S;S_n)$ still satisfies the
absorbing barrier boundary condition
$\Pi_{\eps=0}(\delta_0;\delta_c;S_n)=0$, even when we include the markovian corrections to
first order. 
In Fig.~\ref{fig:Pimemmem2} we compare
$\Pi_{\eps=0}^{\rm mem}+
\Pi_{\eps=0}^{\rm mem-mem}$ to the zeroth-order term
(\ref{Pix0}).

\subsection{The halo mass function}

We can now compute the first crossing rate using \eq{firstcrossT}.
Since both $\Pi^{\rm mem}_{\eps=0}$ and
$\Pi^{\rm mem-mem}_{\eps=0}$ have been expressed as a  derivative
with respect to $\delta_n$ in \eqs{Pimempan}{Pimemmempan}, 
the integral over $d\delta_n$ is performed trivially, and we get
\be\label{Fmemzero}
{\cal F}^{\rm mem}(S)=
-\frac{\pa}{\pa S}\int_{-\infty}^{\delta_c}d\delta_n\, 
\Pi^{\rm mem}_{\eps=0}(\delta_0=0;\delta_n;S)
=0\, ,
\ee
\bees
{\cal F}^{\rm mem-mem}(S)&=&
-\frac{\pa}{\pa S}\int_{-\infty}^{\delta_c}d\delta_n\,
\Pi^{\rm mem-mem}_{\eps=0}(\delta_0=0;\delta_n;S)\nn\\
&=&-\frac{\pa}{\pa S}\[ \frac{\kappa \delta_c}{\sqrt{2\pi S}}\, 
\int_0^{S}dS_i\, \frac{1}{S_i}\, e^{-\delta_c^2/(2S_i)}\]\nn\\
&=&-\frac{\kappa \delta_c}{\sqrt{2\pi}}\, 
\frac{\pa}{\pa S}\[
\frac{1}{S^{1/2}}\G\(0,\frac{\delta_c^2}{2S}\)\]\label{Fmemmem}\\
&=&
-\frac{\kappa \delta_c}{\sqrt{2\pi}}\, 
\[ \frac{1}{S^{3/2}} e^{-\delta_c^2/(2S)}-
 \frac{1}{2 S^{3/2}}\G\(0,\frac{\delta_c^2}{2S}\)\]\, .\nn
\ees
where $\G(0,z)$ is the incomplete Gamma function.
Putting together \eqss{Fmarkov}{Fmemzero}{Fmemmem} we  find the
first-crossing rate to first order in the non-markovian corrections,
\be\label{Ffinal}
{\cal F}(S)=\frac{1-\kappa}{\sqrt{2\pi}}\, 
\frac{\delta_c}{S^{3/2}} e^{-\delta_c^2/(2S)}+
\frac{\kappa}{2\sqrt{2\pi}} 
\frac{\delta_c}{S^{3/2}}\G\(0,\frac{\delta_c^2}{2S}\)\, .
\ee
The halo mass function in
this approximation is therefore

\begin{figure}
\includegraphics[width=0.4\textwidth]{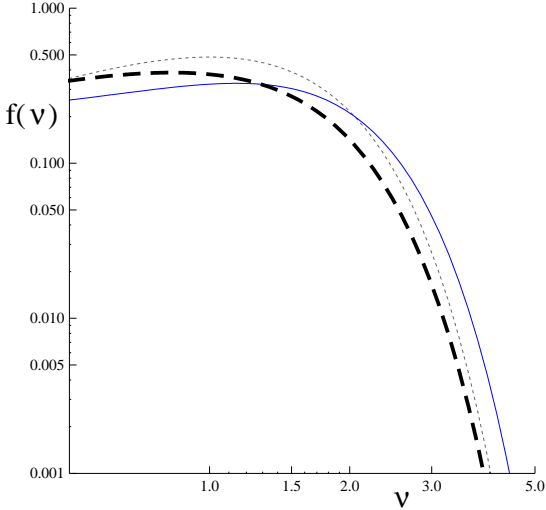}
\caption{\label{fig:compareRKTZ}
The function $f(\nu)$ against $\nu$. The gray dotted line is the
value in PS theory. The blue solid line is the fit to the numerical $N$-body
simulation of \cite{Warren:2005ey}. The thick black dashed line
is our result (\ref{ourf}). 
}
\end{figure}

\be\label{ourf}
f(\s) = (1-\kappa)
\(\frac{2}{\pi}\)^{1/2}\, 
\frac{\d_c}{\s}\, 
\, e^{-\d_c^2/(2\s^2)}
+\frac{\kappa}{\sqrt{2\pi}}\, 
\frac{\d_c}{\s}\, \G\(0,\frac{\d_c^2}{2\s^2}\)\, ,
\ee
where, in the relevant range of values of $R$,
$\kappa$ is given by
\eqs{adiRlim}{Ar}, and is a slowly decreasing function of $R$. For instance,
at $R=5$~{\rm Mpc}, $\kappa \simeq 0.45$, 
at $R=10$~{\rm Mpc}, $\kappa \simeq 0.43$, 
and at $R=20$~{\rm Mpc}, $\kappa \simeq 0.40$., 
For large values of $\d_c^2/2\s^2$
\be
\G\(0,\frac{\d_c^2}{2\s^2}\)\simeq \frac{2\s^2}{\d_c^2}
\, e^{-\d_c^2/(2\s^2)}\, .
\ee
Thus the incomplete Gamma function gives the same exponential factor as
PS theory but with a smaller prefactor,  so
for large halo masses
it is
subleading, and
\eq{ourf} approaches $(1-\kappa)$ times the PS prediction.

In Fig.~\ref{fig:compareRKTZ} we plot the function $f(\nu)$, 
where 
$\nu =\d_c/\s$, 
comparing the prediction of PS theory given in \eq{fps}, 
the fit to the $N$-body simulation
of \cite{Warren:2005ey}, and our result (\ref{ourf}). 
This figure can be compared to 
Fig.~4 of \cite{RKTZ}, see in particular their bottom-left 
panel,
where the authors show  the prediction of PS theory,
the result of their $N$-body simulation, and the computation of 
$ f(\nu)$ with tophat filter in coordinate space, performed with a
Monte Carlo
realization of the trajectories obtained from a
Langevin equation with colored noise. 
We have used the same scale and color code as
their Fig.~4, to make the comparison easier. One sees that our
analytical result for $f(\nu)$ agrees very well with their Monte
Carlo result (the function that we call $f(\nu)$ is
  denoted as $\nu f(\nu)$ in \cite{RKTZ}).
 From \eq{ourf}, we see that in the end our
  expansion parameter is just $\kappa$, so evaluating the
  non-markovian corrections to second order we will get corrections of
  order $\kappa^2$. For $\kappa$ given by \eq{Ar} these are expected
  to be of order $20\%$, which is the level of agreement between our
  analytical result and the Monte Carlo computation.
This provides a non-trivial check of the
correctness of our formalism.

A second consistency check is obtained by recalling
that the fraction of volume occupied by virialized objects 
is given by \eq{F(S)1menoint}.
In hierarchical power spectra, all the mass of the universe must
finally end up in virialized objects, so we must have
$F(S)=1$ when $\d_c/\s\ra 0$. Formally, the limit
$\d_c/\s\ra 0$ can be obtained sending $\d_c\ra 0$ for fixed $\s$, so
we require that
\be\label{checknorma}
\lim_{\d_c\ra 0}\int_{-\infty}^{\d_c} d\d\, \Pi(\d ,S)=0\, .
\ee
As we recalled below
\eq{F(M)PS}, the original PS theory fails this test, giving that only
one half of the total mass of the universe collapses.
In our case
$\Pi=\Pi^{\rm gm}+\Pi^{\rm mem}+\Pi^{\rm mem-mem}$. Since
$\Pi^{\rm gm}$ is the same as in the standard excursion set
result, it already satisfies \eq{checknorma}, 
so we must find that, in the limit
$\d_c\ra 0$, the integral of
$\Pi^{\rm mem}+\Pi^{\rm mem-mem}$ from $-\infty$ to $\d_c$
vanishes. Using 
\eq{Pimem2} we see that
\be
\int_{-\infty}^{\d_c} d\d\,
\Pi_{\eps=0}^{\rm mem}(\d, S)=
\kappa  \[ \frac{\d_c(\d_c-\d)}{S}\, 
{\rm Erfc}\(\frac{2\d_c-\d}{\sqrt{2S}}\)\]_{\d=\d_c}=0\, ,
\ee
for all values of $\d_c$. For the memory-of-memory term
we find
\be
\int_{-\infty}^{\d_c} d\d\,
\Pi_{\eps=0}^{\rm mem-mem}(\d, S)=
\frac{\kappa}{\sqrt{2\pi}}\, 
\frac{\d_c}{S^{1/2}}\G\(0,\frac{\d_c^2}{2S}\)\, .
\ee
Since, for  $z\ra 0$, 
$\G(0,z)\ra -\ln z$, 
we have
\be
\lim_{\d_c\ra 0} \d_c\, \G\(0,\frac{\d_c^2}{2S}\)=0\, ,
\ee
so \eq{checknorma} is indeed satisfied.
An equivalent derivation starts from the observation that, in terms of the function $f(\s)$, the normalization condition reads
\be\label{norma}
\int_0^{\infty}\, \frac{d\s}{\s}f(\s)=1\, .
\ee
Substituting  $f(\s)$ from \eq{ourf} into \eq{norma} and using
\be
\int_0^{\infty}d\s\,   \(\frac{2}{\pi}\)^{1/2}\, 
\frac{\d_c}{\s^2}\, 
\, e^{-\d_c^2/(2\s^2)}=1
\ee
and
\be
\int_0^{\infty}d\s\, \frac{\d_c\,}{\s^2\sqrt{2\pi}}\,
 \G\(0,\frac{ \d_c^2}{2\s^2}\)=1\, ,
 \ee
 we see that the dependence on $\kappa$ cancels and \eq{norma} is satisfied.
The term proportional to the incomplete Gamma function therefore ensure that the mass function is properly normalized, when the amplitude of the term proportional to $\exp\{-\d_c^2/(2\s^2)\}$ is reduced by a factor $1-\kappa$.

A number of comments are now in order. First, our findings
confirms the known result \citep{Bond,RKTZ} that
the corrections obtained by taking properly into account
the tophat filter in coordinate space
do not alleviate the discrepancy of PS theory with the $N$-body
simulations.  We see in fact
from Fig.~\ref{fig:compareRKTZ} that the  
effect of the non-markovian corrections is to give a
halo mass function that, in the relevant mass range, 
is everywhere smaller than the PS mass
function, which results in an improvement in the low-mass range
but in a  worse agreement in the high-mass range. 
This indicates that some crucial physical ingredient is still missing
in the model. This is not surprising at all since, as we already stated, the formation
of dark matter haloes is a complex phenomenon. Incorporating some of the
complexeties within the excursion set theory 
will be the subject of paper~II.

On the positive side, we conclude that we have developed a powerful
analytical formalism that allows us to compute  consistently the
halo mass function when non-markovian
effects are present. In this
paper we have applied it to the corrections generated by the tophat
filter function in coordinate space. However, the same formalism
allows us to compute perturbatively the effect 
of the non-Gaussianities on the halo mass function. This direction 
will be developed in paper~III.

Before leaving this topic we observe that, in the
perturbative computation performed in this section, all terms turned
out to be finite in the continuum limit. The fact that the total
result is finite is obvious for physical reasons. However, the fact that all
the terms that enters in the computation are
separately finite happens to be a happy  accident, related to the
form (\ref{approxDelta}) of $\D(S_i,S_j)$, and in particular  to
the property $\D(S_i,S_i)=0$. However, not all perturbations that we
will consider share this property, in particular when we consider the
non-Gaussianities. Furthermore, even with the above form of 
$\D(S_i,S_j)$, if we work to second order in perturbation theory we
find that divergences appear. It is therefore important to understand
in some detail how in the general case
these divergences cancel among different terms,
giving  a finite result. This issue is quite technical,
and is discussed in detail in 
Appendix~B.

\vspace{5mm}

\noindent We thank Sabino Matarrese
for useful discussions. The work
of MM is supported by the Fond National Suisse. The work of AR is supported by 
the European Community's Research Training Networks under contract MRTN-CT-2006-035505.

\appendix

\section{A. Finite-$\eps$ corrections}\label{app:A}

In this appendix we derive the
results for $\Pi_{\eps}(\delta_0;\delta_c;S)$ and $\Pi_{\eps}(\delta_c;\delta_c;S)$
mentioned in Section~\ref{sect:finite_eps}. 
These results are needed in Section~\ref{sect:nonmark}, when we
compute perturbatively the non-markovian corrections. 
We will also show that, for 
$\pa_{\d}\Pi^{\rm gm}_{\eps}(\delta_0=0;\d;S)$,
the limit $\eps\ra 0^+$ does not commute with the limit $\d\ra \delta_c^-$.
This result will be important in appendix~\ref{app:B}, when we study
the cancellation of divergences that can appear in intermediate steps of
the computation. In \eqs{Pigammaeps}{defgamma} we found that
$\Pi^{\rm gm}_{\eps} (\delta_0;\delta_c;S)=
\sqrt{\eps}\,\, \gamma
(\delta_c/S^{3/2}) e^{-\delta_c^2/(2S)} +{\cal O}(\eps)$,
where
$\gamma = (2/\sqrt{\pi})\,\lim_{\eta\ra 0}\eta\, u(\eta)$.
One possible route to the evaluation of $\gamma$ could be to 
plug \eq{PiCu} into \eq{CK3} and evaluate both sides at $\d=\delta_c$. To
lowest order in $\eps$ one can replace $S+\eps$ on the left-hand side
simply by $S$, and one obtains an integral equation for the unknown 
function $u(\eta)$. This integral equation has the form of a
Wiener-Hopf equation, for which various techniques have been 
developed \citep{Noble}. However, we have found a simpler way to get
directly  $\gamma$, as follows.
We consider the derivative of $\Pi^{\rm gm}_{\eps}$ with
respect to $\delta_c$ (which, when we use the notation
$\Pi^{\rm gm}_{\eps}(\delta_0;\d;S)$,
does not appear explicitly in the list of
variable on which  $\Pi^{\rm gm}_{\eps}$ depends, but of course
enters as upper integration limit in
\eq{defPi}). This gives
\be\label{hatlist}
\frac{\pa}{\pa \delta_c}\Pi_{\eps} (\delta_0;\delta_n;S_n)=
\sum_{i=1}^{n-1}
\int_{-\infty}^{\delta_c} d\delta_1\ldots \widehat{d\d}_i\ldots 
d\delta_{n-1}\,
W(\delta_0;\delta_1,\ldots, \delta_i=\delta_c,\ldots ,\delta_n;S_n)\, ,
\ee
where the notation $ \widehat{d\d}_i$ means that we must omit $d\delta_i$ from
the list of integration variables.
We next use \eqs{facto}{WPiPi} and,
in the continuum limit, we  obtain the identity
\be\label{identity}
\frac{\pa}{\pa \delta_c}\Pi_{\eps=0} (\delta_0;\delta_n;S_n)
=\int_0^{S_n}dS_i\, \lim_{\eps\ra 0}\frac{1}{\eps}
\Pi^{\rm gm}_{\eps}(\delta_0;\delta_c;S_i)
\Pi^{\rm gm}_{\eps}(\delta_c;\delta_n;S_n-S_i)\, .
\ee
The left-hand side of this identity can be evaluated explicitly using
\eq{Pix0} and, setting for simplicity $\delta_0=0$, is
\be
\frac{\pa}{\pa \delta_c}\Pi_{\eps=0} (\delta_0=0;\delta_n;S_n)
=\(\frac{2}{\pi}\)^{1/2}\, \frac{2\delta_c-\delta_n}{S_n^{3/2}}\,
e^{-(2\delta_c-\delta_n)^2/(2S_n)}\, .
\ee
The right-hand side of \eq{identity}
can be evaluated using \eq{Pigammaeps}
together with
\be\label{reshuf}
\Pi_{\eps}(\delta_c;\delta_n;S)=\Pi_{\eps}(\delta_n;\delta_c;S)=
\sqrt{\eps}\, \gamma
\frac{\delta_c-\delta_n}{S^{3/2}} e^{-(\delta_c-\delta_n)^2/(2S)} +{\cal O}(\eps)\, ,
\ee
which can be checked from \eqs{defPi}{W} by performing a reshuffling 
of the dummy integration variables. We see that the limit $\eps\ra 0$
in \eq{identity}
is finite thanks to the factors $\sqrt{\eps}$ in
$\Pi^{\rm gm}_{\eps}(\delta_0;\delta_c;S_i)$ and in
$\Pi^{\rm gm}_{\eps}(\delta_c;\delta_n;S_n-S_i)$. The integral over $S_i$ can
be performed using
the identity
\be\label{magic}
\int_0^{S_n}dS_i\, \frac{1}{S_i^{3/2} (S_n-S_i)^{3/2}}\,
\exp\left\{-\frac{a^2}{2S_i}-\frac{b^2}{2(S_n-S_i)}\right\}
= \sqrt{2\pi}\,\, \frac{a+b}{ab}\, \frac{1}{S_n^{3/2}}
\exp\left\{-\frac{(a+b)^2}{2S_n}\right\}\, ,
\ee
where $a>0, b>0$.\footnote{We have not been able to find this
  identity in standard tables of integrals, but we have verified it
  numerically, with very high accuracy, in a wide range of values of $a$
  and $b$. We can also turn the argument around and say that, since we
know that  $\Pi^{\rm gm}_{\eps}(\delta_0;\delta_c;S)$ has the functional form 
(\ref{Pigammaeps}) and we know that the identity 
(\ref{identity}) holds, it follows that the integral on the left-hand
side of \eq{magic} must be given by the expression on the right-hand
side, times an unknown numerical constant. The latter can  be computed
evaluating the term $\sim 1/a$ of the 
integral in the limit $a\ra 0^+$. 
This is easily done analytically, since in this case the factors
$(S_n-S_i)$ inside the integrand 
can be simply replaced by $S_n$, and fixes the factor
$\sqrt{2\pi}$ on the right-hand side of \eq{magic}.
}
In this way we find that the dependence on $\delta_c$ and $S$ on the two
sides of \eq{identity} is the same, as it should, and we fix
$\gamma =1/\sqrt{\pi}$.

In appendix~\ref{app:B}, when we study the cancellation of divergences,
we will also need 
$\pa_{\d}\Pi^{\rm gm}_{\eps}$, evaluated in $\d=\delta_c$. Of course, 
if we
first take the limit $\eps\ra 0^+$, and then we take $\d\ra \delta_c^-$, we
simply get the derivative of the function
$\Pi_{\eps=0}(\delta_0;\d,S)$ given in \eq{Pix0}, evaluated in $\delta_c$,
\be\label{limlim1}
\lim_{\d\ra \delta_c^-}\, \lim_{\eps\ra 0^+} 
\pa_{\d}\Pi^{\rm gm}_{\eps}(\delta_0=0;\d,S)=
\left.\pa_{\d}\Pi^{\rm gm}_{\eps=0}(\delta_0;\d,S)\right|_{\d=\delta_c}
=
-\(\frac{2}{\pi}\)^{1/2}\frac{\delta_c}{S^{3/2}}\, e^{-\delta_c^2/(2S)}\, .
\ee
However, 
we will actually need the result when the limits are evaluated in the
opposite order, i.e.
$ \lim_{\eps\ra 0^+}\, \lim_{\d\ra \delta_c^-}
\pa_{\d}\Pi^{\rm gm}_{\eps}(\delta_0;\d,S)$.
We will now show that these two limits do not commute. From
\eq{defgamma}, for small $\eta$, $u(\eta)$ is proportional to
$\gamma\sqrt{\pi}/(2\eta)=
1/(2\eta)$. More generally, for small $\eta$, 
we write
\be\label{uu0u1}
u(\eta)=\frac{1}{2\eta}+u_0 +u_1\eta +{\cal O}(\eta^2)\, .
\ee
Plugging this expansion, 
together with the expansion in powers of $\eta$
of \eq{Ceps}, into \eq{PiCu} we find that, for $\eta\ra 0$ (i.e. 
for $\d\ra \delta_c^-$ at fixed $\eps$), retaining only the terms up to
${\cal O}(\sqrt{\eps})$
\be\label{u0etau1eta2}
\Pi^{\rm gm}_{\eps}(\delta_0=0;\d,S)=
\Pi^{\rm gm}_{\eps}(\delta_0=0;\delta_c,S)
+\sqrt{\eps}\, \frac{2}{\sqrt{\pi}}\, (u_0\eta+u_1\eta^2+\ldots)
\frac{\delta_c}{S^{3/2}}\, e^{-\delta_c^2/(2S)}\, .
\ee
Using $\pa_{\d}=(d\eta/d\d)\pa/\pa\eta$ and
$d\eta/d\d=-1/\sqrt{2\eps}$, this gives
\be\label{differsu0}
\lim_{\eps\ra 0^+}\, \lim_{\d\ra \delta_c^-}
\pa_{\d}\Pi^{\rm gm}_{\eps}(\delta_0=0;\d,S)=
-u_0 \(\frac{2}{\pi}\)^{1/2}\frac{\delta_c}{S^{3/2}}\, e^{-\delta_c^2/(2S)}\, ,
\ee
which differs by a factor $u_0$ from \eq{limlim1}. 
It is also interesting to observe, from \eq{u0etau1eta2}, that
also $\pa^2\Pi^{\rm gm}_{\eps}/\pa\eta^2$, evaluated in $\eta =0$, is
proportional to $\sqrt{\eps}$. Since
\be
\frac{\pa^2\Pi^{\rm gm}_{\eps}}{\pa\d^2}=\frac{1}{2\eps}
\frac{\pa^2\Pi^{\rm gm}_{\eps}}{\pa \eta^2}\, ,
\ee
overall $\pa^2\Pi^{\rm gm}_{\eps}/\pa\d^2$, evaluated in $\d=\delta_c$ at
finite $\eps$, is proportional to $1/\sqrt{\eps}$. Therefore, in 
\eq{nogood} the first correction included in the dots, which is proportional  
$\eps(\pa^2\Pi^{\rm gm}_{\eps}/\pa\d^2)_{\d=\delta_c}$, 
is of the same order as the
term
$\sqrt{\eps}(\pa\Pi^{\rm gm}_{\eps}/\pa\d)_{\d=\delta_c}$, and similarly for the
higher-order terms. This is the reason why we could not use 
\eq{nogood} to fix the value of the coefficient $\g$.

Finally, in the perturbative computation we also need
$\Pi^{\rm gm}_{\eps}(\delta_c;\delta_c;S)$, with both arguments equal to
$\delta_c$. The result is given in \eq{Pixcxc}.
To derive it, we first observe from
\eq{Pigammafinal} that, when $\delta_0=\delta_c$, the term ${\cal O}(\sqrt{\eps})$
vanishes, so the first non-vanishing term will be ${\cal O}(\eps)$. 
Invariance under space translations requires
that $\Pi^{\rm gm}_{\eps}(\delta_0;\delta_c;S)$ can depend on $\delta_0$ and
$\delta_c$ only through the combination $\delta_c-\delta_0$, so when $\delta_0=\delta_c$ it becomes a
function of $S$ only. 
We can perform 
dimensional analysis
assigning to $\d$ some (unspecified)
dimension $\ell$ and to $S$ dimensions $\ell^2$. In this case, from
\eq{Langevin2} we see that $\eta$ 
has dimensions $1/\ell$, and $\dot{\xi}\sim \ell/\ell^2=1/\ell$, 
so \eq{Langevin1}
is dimensionally correct. 
In these units $\lambda\sim 1/\ell$, since $\lambda \d$ is
dimensionless, and we see from \eq{Piexplicit} that $\Pi_{\eps}$
has dimensions $1/\ell$. Using dimensional analysis in this form
we conclude that the term ${\cal O}(\eps)$ in 
$\Pi^{\rm gm}_{\eps}(\delta_c;\delta_c;S)$ is necessarily proportional to 
$\eps/S^{3/2}$. Since this fixes completely the dependence on $S$,
writing $S=\eps n$ we have also fixed completely the dependence on
$n$, i.e. to ${\cal O}(\eps)$ we must have
\be\label{Pixcxc2}
\Pi^{\rm gm}_{\eps}(\delta_c;\delta_c;S)=
c\, \frac{\eps}{S^{3/2}}=
\frac{c}{\sqrt{\eps}\, n^{3/2}}
\, ,
\ee
with $c$ independent of $n$. 
The coefficient $c$  can then be fixed
computing explicitly the integral 
in \eq{defPi} when $n=2$, i.e. when there is just one integration
variable. This can be done analytically and shows that
$c=1/\sqrt{2\pi}$. The computation for $n=2$ actually shows that
\eq{Pixcxc2} is  exact, i.e. it receives no correction of higher
order in $\eps$. Even for $n=3$ the integral in
\eq{Pixcxc2} can be performed analytically when $\delta_0=\delta_c$, and again
we find that \eq{Pixcxc2} is exact. We have checked this result
numerically for $n$ up to 7 and we find that the numerical result
agrees with \eq{Pixcxc2} within the 10 digit precision of the
numerical integration, so it is clear that \eq{Pixcxc2} is actually
exact, and not just the result at ${\cal O}(\eps)$. (In any case, to perform
our perturbative
computation, we only need $\Pi^{\rm gm}_{\eps}(\delta_c;\delta_c;S)$ to
${\cal O}(\eps)$.)

\section{B. Divergences and the finite part prescription}\label{app:B}

In this appendix we first of all
reconsider the perturbative computation of 
Section~\ref{sect:nonmark} for a generic function $\D_{ij}$
(still symmetric in $(i,j)$). This will reveal some complexities that
were not apparent in the computation of Section~\ref{sect:nonmark},
and that will be very important when computing the
non-Gaussianities. 
If $\D_{ij}$ does not vanish when $i=j$, we rewrite \eq{Delta1} as
\be\label{Delta1app}
\Pi_{\eps}(\delta_0;\delta_n;S_n) =
\int_{-\infty}^{\delta_c} 
d\delta_1\ldots d\delta_{n-1}\,\Dl
\exp\left\{\frac{1}{2}\sum_{i,j=1}^n\D_{ij}\pa_i\pa_j
\right\}\exp\left\{
i\sum_{i=1}^n\lambda_i\delta_i -\frac{1}{2}\sum_{i,j=1}^n
[{\rm min}(S_i,S_j)]
\lambda_i\lambda_j\right\}\, ,
\ee
where, as usual, we used the fact that, acting on $\exp\{i\lambda_i \delta_i\}$, 
$\pa_i$ gives $i\lambda_i$.
Since $\D_{nn}$ is now in general non-vanishing, in the
sum (\ref{Dsum}) the term $\D_{nn}\pa_n^2$ 
contributes. Furthermore, now
\be
\frac{1}{2}\sum_{i,j=1}^{n-1}\D_{ij}\pa_i\pa_j=
\sum_{i<j}\D_{ij}\pa_i\pa_j +
\frac{1}{2}\sum_{i}^{n-1}\D_{ii}\pa^2_i\, .
\ee
The operator $\exp\{(1/2)\D_{nn}\pa_n^2\}$ can be  carried out of
the integral over $d\delta_1, \ldots ,d\delta_{n-1}$, 
while the other terms $\D_{ij}$ 
will be expanded perturbatively. Thus, \eqst{PiD1}{defPimem-memory}
are replaced by
\be\label{PiDnn}
\Pi^{{\D}1}_{\eps}=e^{(1/2)\D_{nn}\pa_n^2}
\[\Pi_{\eps}^{\rm mem}+\Pi_{\eps}^{\rm mem-mem}\]\, ,
\ee
where
\be\label{defPimemoryapp}
\Pi_{\eps}^{\rm mem}(\delta_0;\delta_n;S_n)
=\sum_{i=1}^{n-1}\D_{in}\pa_n
\int_{-\infty}^{\delta_c} 
d\delta_1\ldots d\delta_{n-1}\,\pa_iW^{\rm gm}(\delta_0;\delta_1,\ldots ,\delta_n;S_n)\, ,
\ee
and
\be\label{defPimem-memoryapp}
\Pi_{\eps}^{\rm mem-mem}(\delta_0;\delta_n;S_n)=
\int_{-\infty}^{\delta_c} 
d\delta_1\ldots d\delta_{n-1} 
\[\sum_{i<j}\D_{ij}\pa_i\pa_j
+\frac{1}{2}\sum_{i}^{n-1}\D_{ii}\pa_i^2\]
W^{\rm gm}(\delta_0;\delta_1,\ldots ,\delta_n;S_n)
\, .\nn
\ee
The memory term is the same as in Section~\ref{sect:nonmark}, so it is
still finite. 
The memory-of-memory term, however, presents a new difficulty.
Using \eq{facto}
we get
\bees
\Pi_{\eps}^{\rm mem-mem}(\delta_0;\delta_n;S_n)
&=&\sum_{i<j}\D_{ij}
\Pi^{\rm gm}_{\eps}(\delta_0;\delta_c;S_i)
\Pi^{\rm gm}_{\eps}(\delta_c;\delta_c;S_j-S_i)
\Pi^{\rm gm}_{\eps}(\delta_c;\delta_n;S_n-S_j)\nn\\
&&+
\sum_{i=1}^{n-1}\frac{\D_{ii}}{2}
\pa_i\[ \Pi^{\rm gm}_{\eps}(\delta_0;\delta_i;S_i)
\Pi^{\rm gm}_{\eps}(\delta_i;\delta_n;S_n-S_i)\]_{\delta_i=\delta_c}\label{Pimemmem1app}\\
&&\hspace*{-8mm}=\sum_{i<j}\D_{ij}
\Pi^{\rm gm}_{\eps}(\delta_0;\delta_c;S_i)
\Pi^{\rm gm}_{\eps}(\delta_c;\delta_c;S_j-S_i)
\Pi^{\rm gm}_{\eps}(\delta_c;\delta_n;S_n-S_j)
+\sum_{i=1}^{n-1}\D_{ii}
[\pa_i\Pi^{\rm gm}_{\eps}(\delta_0;\delta_i;S_i)]_{\delta_i=\delta_c}
\Pi^{\rm gm}_{\eps}(\delta_c;\delta_n;S_n-S_i)\, .\nn
\ees
We now discover  that the continuum limit of the
memory-of-memory term is non-trivial, since it is made of two terms
that  are separately
divergent. Consider first the second  term 
in \eq{Pimemmem1app}, which is the one 
coming from
$\D_{ii}\pa_i^2$.
We have found in
Section~\ref{sect:finite_eps} that 
$\Pi^{\rm  gau}_{\eps}(\delta_0;\delta_c;S_i)$ is proportional to
$\sqrt{\eps}$ while
$[\pa_{\d}\Pi^{\rm gm}_{\eps}(\delta_c;S;S)]_{\d=\delta_c}$ has a finite 
limit for $\eps\ra 0$, see \eq{differsu0}.
Therefore, using \eq{sumint}, we find
that the last term in \eq{Pimemmem1app} diverges as $1/\sqrt{\eps}$.
A similar problem appears in the  term coming from $\pa_i\pa_j$ with
$i\neq j$. Using
\eqs{Pigammafinal}{Pixcxc} we find
that the first
term in \eq{Pimemmem1app} is proportional to
\be\label{diver1}
\eps \sum_{i=1}^{n-2}\frac{1}{S_i^{3/2} }
\exp\left\{-\frac{\delta_c^2}{2S_i}\right\}
 \eps\sum_{j=i+1}^{n-1}
\frac{\D_{ij}}{(S_j-S_i)^{3/2} (S_n-S_j)^{3/2}}
\exp\left\{-\frac{(\delta_c-\delta_n)^2}{2(S_n-S_j)}\right\}\, ,
\ee
where $S_i=i\eps, S_j=j\eps$. 
In the continuum limit, unless $\D_{ij}$ vanishes for $i=j$,
this quantity diverges as $1/\sqrt{\eps}$,
because of the  behavior  $(S_j-S_i)^{-3/2}$ when
$S_j\ra S_i^+$. In Section~\ref{sect:finite_eps} these problem
did not show up because $\D_{ii}=0$, so the divergence coming from
$\D_{ii}\pa_i^2$ disappears. Furthermore, when $S_i\ra S_j$,
$\D_{ij}$ vanished as $S_j-S_i$, thereby ensuring the convergence of
the sum (or, in the continuum limit, of the integral 
over $S_j$)
in \eq{diver1}.

In order to understand how the cancellation mechanism works when
$\D_{ij}$ does not vanish for $S_i=S_j$,
we  examine the memory-of-memory term when $\D(S_i,S_j)$ is a
constant, that we set equal to unity.  The reason is that, in this case,
we can compute it  in an alternative way, which  shows
that the result is finite.
The trick is to compute the second derivative of
$\Pi^{\rm gm}_{\eps}$ with respect to $\delta_c$. The first derivative
was computed in \eq{hatlist}, and the result can be rewritten as
\be
\frac{\pa}{\pa \delta_c}\Pi^{\rm gm}_{\eps} (\delta_0;\delta_n;S_n)=
\sum_{i=1}^{n-1}
\int_{-\infty}^{\delta_c} d\delta_1\ldots d\delta_{n-1}\, 
\pa_iW\, .
\ee
When we take one more derivative of \eq{hatlist} with respect to
$\delta_c$, we find two kinds of terms. First, there are the terms where we
take one more derivatives with respect to the upper limit of the integration
with respect to a variable $d\delta_j$ with $j\neq i$. Furthermore, we must
take the derivative of $W(\delta_0;\delta_1,\ldots, \delta_i=\delta_c,\ldots
,\delta_{n-1},\delta_n;S_n)$ with respect to $\delta_c$. Therefore
\bees
\frac{\pa^2}{\pa \delta_c^2}\Pi^{\rm gm}_{\eps} (\delta_0;\delta_n;S_n)&=&
2\sum_{ i<j }
\int_{-\infty}^{\delta_c} d\delta_1\ldots \widehat{d\d}_i\ldots 
\widehat{d\d}_j\ldots d\delta_{n-1}\,
W(\delta_0;\delta_1,\ldots, \delta_i=\delta_c,\ldots ,\delta_j=\delta_c,\ldots ,\delta_n;S_n)
\nn\\
&&+\sum_{i=1}^{n-1}\int_{-\infty}^{\delta_c} d\delta_1\ldots \widehat{d\d}_i\ldots 
 d\delta_{n-1}\,
\frac{\pa}{\pa \delta_c}
W(\delta_0;\delta_1,\ldots, \delta_i=\delta_c,\ldots ,\delta_n;S_n)\nn\\
&=&2\sum_{i<j}
\int_{-\infty}^{\delta_c} d\delta_1\ldots d\delta_{n-1}\pa_i\pa_jW
+\sum_{i=1}^{n-1}
\int_{-\infty}^{\delta_c} d\delta_1\ldots d\delta_{n-1}\pa^2_iW\, ,\label{hatlist2}
\ees
that is,
\be
\frac{\pa^2}{\pa \delta_c^2}\Pi^{\rm gm}_{\eps} (\delta_0;\delta_n;S_n)=
\sum_{i,j=1}^{n-1}
\int_{-\infty}^{\delta_c} d\delta_1\ldots d\delta_{n-1}\pa_i\pa_jW\, .
\ee
Thus, when $\D_{ij}=1$,
\be
\Pi_{\eps}^{\rm mem-mem}(\delta_0;\delta_n;S_n)=
\frac{1}{2}\frac{\pa^2}{\pa \delta_c^2}\Pi^{\rm gm}_{\eps} (\delta_0;\delta_n;S_n)\, .
\ee
In particular, in the continuum limit,
\be\label{bench}
\Pi_{\eps=0}^{\rm mem-mem}(\delta_0=0;\delta_n;S_n)=
\frac{1}{2}\frac{\pa^2}{\pa \delta_c^2}\Pi^{\rm gm}_{\eps=0}
(\delta_0=0;\delta_n;S_n)
=\(\frac{2}{\pi}\)^{1/2}
\[ 1-\frac{(2\delta_c-\delta_n)^2}{S_n}\]\, \frac{1}{S_n^{3/2}}
e^{-(2\delta_c-\delta_n)^2/(2S_n)}\, .
\ee
First of all this result shows that, when $\D_{ij}=1$,
$\Pi_{\eps}^{\rm mem-mem}$ stays indeed
finite in the continuum limit. 
Second, it gives its explicit expression, which can then be
compared with a computation based on \eq{Pimemmem1app}.
To perform the comparison, we first compute 
the second term in \eq{Pimemmem1app}, when
$\D_{ij}=1$, i.e.
\be
I_1\equiv \sum_{i=1}^{n-1}
[\pa_i\Pi^{\rm gm}_{\eps}(\delta_0=0;\delta_i;S_i)]_{\delta_i=\delta_c}
\Pi^{\rm gm}_{\eps}(\delta_c;\delta_n;S_n-S_i)\, .
\ee
Observe that  in this expression we must first compute the
derivative in $\delta_i=\delta_c$ (since this came from the integration by parts
of $\pa_i^2$) and only after we take the limit $\eps\ra 0^+$. The
result is therefore given by \eq{differsu0}. Using also 
\eqs{reshuf}{Pigammafinal}, we get
\be
I_1=-\frac{1}{\sqrt{\eps}}
\,\frac{u_0\sqrt{2}}{\pi}\delta_c(\delta_c-\delta_n)\eps \sum_{i=1}^{n-1}
\frac{1}{S_i^{3/2}(S_n-S_i)^{3/2}}\,
\exp\left\{-\frac{\delta_c^2}{2S_i}
-\frac{(\delta_c-\delta_n)^2}{2(S_n-S_i)}\right\}\, .
\ee
Because of the exponential factor, the argument of the sum goes to
zero very fast as $S_i\ra 0^+$ and as $S_i\ra S_n^-$, and
therefore we can use \eq{sumint}, so
\be
I_1=-\frac{1}{\sqrt{\eps}}
\,\frac{u_0\sqrt{2}}{\pi}\delta_c(\delta_c-\delta_n)\int_{0}^{S_n}dS_i\, 
\frac{1}{S_i^{3/2}(S_n-S_i)^{3/2}}\,
\exp\left\{-\frac{\delta_c^2}{2S_i}
-\frac{(\delta_c-\delta_n)^2}{2(S_n-S_i)}\right\}\, .
\ee
The integral can be performed using \eq{magic}, and we  get
\be\label{I1}
I_1
=-\frac{1}{\sqrt{\eps}}
\,\frac{2u_0}{\sqrt{\pi}} (2\delta_c-\delta_n)
\frac{1}{S_n^{3/2}}
e^{-(2\delta_c-\delta_n)^2/(2S_n)}\, .
\ee
Therefore $I_1$ diverges as $1/\sqrt{\eps}$. It is important to
observe that there is no finite part in $I_1$. In the continuum limit
the corrections
to \eqss{Pigammafinal}{differsu0}{sumint} 
are all ${\cal O}(\eps)$ compared to the
leading terms that we used, so they produce terms that are overall
${\cal O}(\sqrt{\eps})$ in \eq{I1}, and therefore vanish in the continuum limit.

We next consider the other term
in \eq{Pimemmem1app}, i.e.
\bees
I_2&\equiv& \sum_{i=1}^{n-2}
\Pi^{\rm gm}_{\eps}(\delta_0=0;\delta_c;S_i)
\sum_{j=i+1}^{n-1}
\Pi^{\rm gm}_{\eps}(\delta_c;\delta_c;S_j-S_i)
\Pi^{\rm gm}_{\eps}(\delta_c;\delta_n;S_n-S_j)\nn\\
&=&\frac{1}{\pi\sqrt{2\pi}}\, \delta_c(\delta_c-\delta_n)\, 
\eps\sum_{i=1}^{n-2}\,\frac{1}{S_i^{3/2}}
e^{-\delta_c^2/(2S_i)}
\eps \sum_{j=i+1}^{n-1}
\frac{1}{(S_j-S_i)^{3/2}(S_n-S_j)^{3/2}}\,
\exp\left\{
-\frac{(\delta_c-\delta_n)^2}{2(S_n-S_j)}\right\}\, .\label{defI2}
\ees
Now the passage from the sums to integrals is more delicate. One might
be tempted to write
\be\label{sumint2}
\eps \sum_{j=i+1}^{n-1} \stackrel{?}{=}
\int_{S_i}^{S_n}dS_j\, .
\ee
However, \eq{sumint2}
is only correct when
the sum and the integral are finite for $\eps\ra 0$. Here this is not
the case, since
\be
\int_{S_i}^{S_n}dS_j\, 
\frac{1}{(S_j-S_i)^{3/2}(S_n-S_j)^{3/2}}\,
\exp\left\{
-\frac{(\delta_c-\delta_n)^2}{2(S_n-S_j)}\right\}
\ee
diverges at the lower integration limit $S_j=S_i$, and indeed
our aim is to extract this divergent term, plus the finite terms.
A  better guess would be that, since the sum starts from $j=i+1$, the
corresponding integral should start from $S_j=S_i+\eps$, so
\be\label{sumint3}\label{I3presc}
I_3\equiv \eps \sum_{j=i+1}^{n-1}
\frac{1}{(S_j-S_i)^{3/2}(S_n-S_j)^{3/2}}\,
\exp\left\{
-\frac{(\delta_c-\delta_n)^2}{2(S_n-S_j)}\right\}\stackrel{?}{=} 
\int_{S_i+\eps}^{S_n}dS_j
\frac{1}{(S_j-S_i)^{3/2}(S_n-S_j)^{3/2}}\,
\exp\left\{
-\frac{(\delta_c-\delta_n)^2}{2(S_n-S_j)}\right\}
\, .
\ee
Still, this  cannot be completely correct. To realize this observe
that, since the integral is dominated by $S_j=S_i+\eps$, the
divergent part can be extracted replacing $S_j=S_i$ everywhere
except in the factor $(S_j-S_i)^{-3/2}$ so,
if we used this
prescription, we would
conclude that 
\bees
I_3&\stackrel{?}{=}&\frac{1}{(S_n-S_i)^{3/2}}\,
\exp\left\{
-\frac{(\delta_c-\delta_n)^2}{2(S_n-S_i)}\right\}
\int_{S_i+\eps}dS_j
\frac{1}{(S_j-S_i)^{3/2}} +\,\, {\rm finite\,\,  parts}\nn\\
&=&\frac{2}{\sqrt{\eps}} \frac{1}{(S_n-S_i)^{3/2}}\,
\exp\left\{
-\frac{(\delta_c-\delta_n)^2}{2(S_n-S_i)}\right\}
+{\rm finite\,\,  parts}\, .
\ees
However, if the prescription (\ref{I3presc}) where correct in general,
we should
 get the same result if we separate the term $j=i+1$ from the sum,
and we let the remaining integral start from $S_j=S_i+2\eps$, 
so we should get the same result if we write
\bees
I_3 &\stackrel{?}{=}& 
\frac{1}{\sqrt{\eps}}\frac{1}{(S_n-S_i)^{3/2}}\,
\exp\left\{
-\frac{(\delta_c-\delta_n)^2}{2(S_n-S_i)}\right\}
+\int_{S_i+2\eps}^{S_n}dS_j
\frac{1}{(S_j-S_i)^{3/2}(S_n-S_j)^{3/2}}\,
\exp\left\{
-\frac{(\delta_c-\delta_n)^2}{2(S_n-S_j)}\right\}\nn\\
&=&\frac{1+\sqrt{2}}{\sqrt{\eps}} \frac{1}{(S_n-S_i)^{3/2}}\,
\exp\left\{
-\frac{(\delta_c-\delta_n)^2}{2(S_n-S_i)}\right\}
+{\rm finite\,\,  parts}\, .
\ees
We see that the two procedures both agree on the fact that the
singularity is proportional to $1/\sqrt{\eps}$, but give different
values for the coefficient, so \eq{I3presc} cannot correct in general.
Observe also that the finite parts are not affected by this
ambiguity, which amounts to a rescaling of $\eps$.

Since, of course, the strength of the singularity is in principle
fixed (although difficult to compute analytically)
as long as we write $I_3$ as a sum,  we can always
choose a value $\a$ such that, as far as  the $1/\sqrt{\eps}$
singularity and the finite terms are concerned, we have the equality
\be\label{I3presc2}
\eps\sum_{j=i+1}^{n-1}
\frac{1}{(S_j-S_i)^{3/2}(S_n-S_j)^{3/2}}\,
\exp\left\{
-\frac{(\delta_c-\delta_n)^2}{2(S_n-S_j)}\right\}
=
\int_{S_i+\a\eps}^{S_n}dS_j
\frac{1}{(S_j-S_i)^{3/2}(S_n-S_j)^{3/2}}\,
\exp\left\{
-\frac{(\delta_c-\delta_n)^2}{2(S_n-S_j)}\right\}
\, ,
\ee
and the two expressions only differ by terms that vanish as $\eps\ra 0$.
In fact, $\a$ can be fixed requiring that the coefficient of
$1/\sqrt{\eps}$ is the same on the two sides of \eq{I3presc2}, and it
does not affect the terms ${\cal O}(\eps^0)$ since it it just a rescaling of
$\eps$.  Actually, in our problem, an even better way to pass from the
sum to the integral is to write
\be\label{I3presc3}
\eps\sum_{j=i+1}^{n-1}
\frac{1}{(S_j-S_i)^{3/2}(S_n-S_j)^{3/2}}\,
\exp\left\{
-\frac{(\delta_c-\delta_n)^2}{2(S_n-S_j)}\right\}
=\int_{S_i}^{S_n}dS_j
\frac{1}{(S_j-S_i)^{3/2}(S_n-S_j)^{3/2}}\,
\exp\left\{-\frac{\a\eps}{2(S_j-S_i)}
-\frac{(\delta_c-\delta_n)^2}{2(S_n-S_j)}\right\}
\, .
\ee
In other words, rather than setting the integrand to zero for 
$S_j<S_i+\a\eps$, we cut it off exponentially
using the factor $\exp\{-\a\eps/(S_j-S_i)\}$. Again this produces
a $1/\sqrt{\a\eps}$ singularity, as we will check in a moment, and $\a$
can be chosen so that this singularity has the same strength
as that on the
left-hand side of \eq{I3presc3}. However, since $\a$ is just
a rescaling of
$\eps$,  $\a$ does not affect the finite part.

The advantage of using \eq{I3presc3} is that the integral can now be
performed analytically using
\eq{magic}, so we get
\be
I_3= \eps\sum_{j=i+1}^{n-1}
\frac{1}{(S_j-S_i)^{3/2}(S_n-S_j)^{3/2}}\,
\exp\left\{
-\frac{(\delta_c-\delta_n)^2}{2(S_n-S_j)}\right\}
=
\sqrt{2\pi}\,\(\frac{1}{\sqrt{\a\eps}} +\frac{1}{\delta_c-\delta_n}\)
\frac{1}{(S_n-S_i)^{3/2}}\, 
\exp\left\{-\frac{(\delta_c-\delta_n+\sqrt{\a\eps})^2}{2(S_n-S_i)}\right\}
\, .\label{I3presc4}
\ee
We have also checked this result numerically. The sum on the left-hand side
can be computed very easily numerically, say for $n$ up to $10^4$, 
and we find that the  right-hand side
reproduces it perfectly, for all values of $\delta_n$, $S_i$ and $S_n$,
if we choose $\a\simeq 0.92$.
Expanding the dependence on $\a\eps$ in the exponential and omitting
the terms that vanish in the limit $\eps\ra 0$, we 
find
\be
I_3=
\sqrt{2\pi}
\[\frac{1}{\sqrt{\a\eps}}+\(1-\frac{(\delta_c-\delta_n)^2}{S_n-S_i}\)
\frac{1}{\delta_c-\delta_n}\] 
\frac{1}{(S_n-S_i)^{3/2}}\,
\exp\left\{-\frac{(\delta_c-\delta_n)^2}{2(S_n-S_i)}\right\}
\, ,
\ee
which explicitly displays the $1/\sqrt{\eps}$ singularity and the
finite, $\a$-independent, part.

To compute $I_2$ we must still plug this expression into \eq{defI2}
and carry out the sum over $i$. The latter sum presents no difficulty
since its argument converges well both at $S_i=0$ and at
$S_i=S_n$, so we can just replace the sum by an integral using
\eq{sumint}. It is actually convenient to leave $I_3$ in the form
\eq{I3presc4}, so the integral over $S_i$ can again be performed
using \eq{magic}, and we finally get
\be
I_2= \(\frac{2}{\pi}\)^{1/2} \frac{1}{S_n^{3/2}}
e^{-(2\delta_c-\delta_n)^2/(2S_n)} \[ \frac{2\delta_c-\delta_n}{\sqrt{\a\eps}}
+\( 1-\frac{(2\delta_c-\delta_n)^2}{S_n}\)\]\, .
\ee
Putting together this result and \eq{I1} we finally find
\be
\Pi_{\eps=0}^{\rm mem-mem}(\delta_0;\delta_n;S_n)
=\frac{1}{\sqrt{\eps}}
\( \frac{1}{\sqrt{\a}}  -u_0\sqrt{2} \) (2\delta_c-\delta_n)
\frac{1}{S_n^{3/2}}
e^{-(2\delta_c-\delta_n)^2/(2S_n)}+\(\frac{2}{\pi}\)^{1/2}
\[ 1-\frac{(2\delta_c-\delta_n)^2}{S_n}\]
 \frac{1}{S_n^{3/2}}
e^{-(2\delta_c-\delta_n)^2/(2S_n)}
\, .\label{bench2}
\ee
However, in this case we already know the exact result for
$\Pi_{\eps=0}^{\rm mem-mem}$, which is given by
\eq{bench}. Comparing these two results we learn the following.
First, we know from \eq{bench} that the result is finite and there is
no $1/\sqrt{\eps}$ term. In the computation leading to 
\eq{bench2} we rather find two separately divergent contribution, so they
must cancel.
This is fully consistent with 
\eq{bench2}, since
these divergent terms have exactly the same
dependence on $\eps$, $\delta_c$, $\delta_n$ and $S_n$. We also see 
that, in this second way of performing the computation, the
cancellation depends on the numerical values of quantities, such
as $u_0$, that
are determined by the solution in the boundary layer, and which therefore
are  difficult to compute, as well as on the constant $\a$
that we determined numerically.
The finite part is instead completely fixed, and it is not affected by 
the solution in the boundary layer, nor by the constant $\a$, and
correctly reproduces \eq{bench}.

From this explicit example we can now extract a general rule of
computation. Whenever $\D(S_i,S_j)$ is a regular function, 
such as that given in \eq{approxDelta}, 
the memory-of-memory term and analogous quantities which are finite
when $\D(S_i,S_j)=1$,
will still be finite. The explicit computation with the formalism
developed in Section~\ref{sect:nonmark} can generate terms that
are separately divergent when $\eps\ra 0^+$. However, since 
the total result is finite,  these
divergences must cancel among them.
When we find integrals that diverge in the limit in which two
integration variables become equal (such as the limit $S_j\ra
S_i$ above)
we can just regularize them as 
in \eq{I3presc3}. We  call this technique ``the
$\a$-regularization''. We then discard the divergence and
we extract the finite part, which  is independent of $\a$.
We will indicate by the symbol ${\cal FP}$ this procedure of taking
the finite part. In this notation, the result of the
above computations can be
summarized by
\bees
{\cal FP}\sum_{i=1}^{n-1}
\int_{-\infty}^{\delta_c} d\delta_1\ldots d\delta_{n-1}\pa^2_iW&=&0\, ,\\
{\cal FP} \sum_{i=1}^{n-2}\sum_{j=i+1}^{n-1}
\int_{-\infty}^{\delta_c} d\delta_1\ldots d\delta_{n-1}\pa_i\pa_jW
&=&\(\frac{2}{\pi}\)^{1/2}
\[ 1-\frac{(2\delta_c-\delta_n)^2}{S_n}\]
 \frac{1}{S_n^{3/2}}
e^{-(2\delta_c-\delta_n)^2/(2S_n)}\, .
\ees
As an application of the above formalism, we have studied what happens
choosing a different expansion point when computing the halo mass
function with a tophat filter in coordinate space.
Observe in fact that, since $\D(S_i,S_j)$ is
symmetric under exchange of $S_i$ with $S_j$, 
\eq{approxDelta}, which is valid for $S_i\leq S_j$, can be rewritten
more generally as
\be
\D(S_i,S_j)\simeq \kappa\, 
 \[ {\rm min}(S_i,S_j) -
\frac{[{\rm min}(S_i,S_j)]^2}{{\rm max}(S_i,S_j)} \]
\, .\label{CTTaTbis}
\ee
Thus, the two-point correlator can be written as
\be\label{rever2}
\langle\delta_i\delta_j\rangle = 
(1+\kappa) {\rm min}(S_i,S_j) + \tilde{\D}(S_i,S_j)\, .
\ee
where,  for $S_i\leq S_j$, $\tilde{\D}(S_i,S_j)=-\kappa S_i^2/S_j$.
We can therefore use
$(1+\kappa) {\rm min}(S_i,S_j)$ as the unperturbed two-point function, 
and treat $\tilde{\D}_{ij}$ as the perturbation. The zeroth order term
can again be computed exactly, since it just amounts to a rescaling of
$S$, $S\ra (1+\kappa) S$. At first sight this seems to give a modified
exponential in the distribution function, since  factors such as
$\exp\{-\delta_c^2/(2S)\}$ in \eq{Pix0} becomes
$\exp\{-\delta_c^2/[2(1+\kappa)S]\}$. However, 
now $\tilde{\D}_{nn}=-\kappa S_n$ is non-zero, and
we should not
forget the factor $\exp\{(1/2)\tilde{\D}_{nn}\pa_n^2\}$ in 
\eq{PiDnn}. The effect of this term  can be computed exactly using the 
identity
\be\label{Pi0gaumu2exact}
\exp\left\{\frac{1}{2} (b-a)\pa_x^2\right\}
\frac{1}{\sqrt{a}}\, e^{-x^2/(2a)}
=\frac{1}{\sqrt{b}}\, e^{-x^2/(2b)}\, ,
\ee
which is valid for $a>0$ and $b> 0$. To prove it, we 
write
\bees
\exp\left\{\frac{1}{2} (b-a)\pa_x^2\right\}
\frac{1}{\sqrt{a}}\, e^{-x^2/(2a)}
&=&
\exp\left\{\frac{1}{2} (b-a)\pa_x^2\right\}
\inT\frac{d\lambda}{2\pi}\, e^{i\lambda x-(1/2) a\lambda^2}
=\inT\frac{d\lambda}{2\pi}\,
\sum_{n=0}^{\infty}\,\frac{1}{n!}
 \(\frac{b-a}{2}\)^n\pa_x^{2n}e^{i\lambda x-(1/2) a\lambda^2}\nn\\
&=&\inT\frac{d\lambda}{2\pi}\,
\sum_{n=0}^{\infty}\,\frac{1}{n!}
 \(\frac{b-a}{2}\)^n(i\lambda)^{2n} e^{i\lambda x-(1/2) a\lambda^2}
=\inT\frac{d\lambda}{2\pi}\, e^{-(1/2) (b-a)\lambda^2}
 e^{i\lambda x-(1/2) a^2\lambda^2 }\nn\\
&=&\inT\frac{d\lambda}{2\pi}\,  e^{i\lambda x-(1/2) b\lambda^2 }=
\frac{1}{\sqrt{b}}\, e^{-x^2/(2b)}
\, .\label{proveid}
\ees
(Observe that for $b<0$ the final integral over $d\lambda$ does not
converge, so this identity only holds if $b>0$). In this way, we find
that the action of $\exp\{(1/2)\tilde{\D}_{nn}\pa_n^2\}$ on
$\exp\{-\delta_c^2/[2(1+\kappa)S]\}$
gives back the ``unperturbed'' exponential
factor $\exp\{-\delta_c^2/(2S)\}$, so the zeroth-order term of this expansion
is finally the same as \eq{Pix0}. The computation of the 
non-markovian corrections  requires the finite part prescription,
since now $\tilde{\D}(S_i,S_j)$ does not vanish for $S_i=S_j$. 
The integrals over $dS_i$ and $dS_j$  are more difficult to
compute, but for $\delta_c^2/(2S)\gg 1$
their exponential dependence is easily computed and,
after taking into
again the operator $\exp\{(1/2)\tilde{\D}_{nn}\pa_n^2\}$ in 
\eq{PiDnn}, we find that the exponential dependence of the corrections
is the same that we obtained in \eq{Ffinal}.

\end{document}